\documentclass[journal,10pt,twocolumn,twoside]{IEEEtran}
\usepackage{multicol}   
\usepackage{multirow}
\usepackage[pdftex]{graphicx}
\usepackage{algorithm,algpseudocode}
\usepackage[small]{caption}
\usepackage{float}
\usepackage{amssymb,amsmath}
\usepackage{graphicx}
\usepackage{subfigure}
\usepackage{xspace}
\usepackage{amsthm}
\usepackage{fancyhdr}

\usepackage{caption,setspace}
\usepackage{amssymb,amsmath}
\usepackage{bm}
\usepackage{textcomp} 
\usepackage{epstopdf}
\usepackage{bbding}
\usepackage{cite} 
\usepackage{color}
\usepackage{pifont}
\usepackage{lipsum}
\usepackage{amstext}
\usepackage{orcidlink}

\usepackage{cuted}
\usepackage{stfloats}

\graphicspath{{fig/}}

\newcommand\scalemath[2]{\scalebox{#1}{\mbox{\ensuremath{\displaystyle #2}}}}

\begin{document}

\title{\huge Robust Transmission Design for RIS-Assisted RSMA-SWIPT Systems With Movable Antennas Under Hardware Distortions}
\author{Muhammad Asif \,\orcidlink{0000-0002-9699-1675}, \IEEEmembership{Member, IEEE}, Asim Ihsan \,\orcidlink{0000-0001-7491-7178}, Irfan Muhammad \,\orcidlink{0000-0001-5703-7988}, \IEEEmembership{Member, IEEE}, Mohd Hamza Naim Shaikh  \,\orcidlink{0000-0002-1869-0009}, \IEEEmembership{Member, IEEE}, Syed Tariq Shah \,\orcidlink{0000-0003-4722-1786}, \IEEEmembership{Member, IEEE}, Zhu Shoujin \,\orcidlink{0000-0002-2615-2489}, and Symeon Chatzinotas \,\orcidlink{0000-0001-5122-0001}, \IEEEmembership{Fellow, IEEE}

\thanks{This work was supported by the Anhui Provincial Higher Education Institutions Young Core Faculty Domestic Visiting Scholar Training Funding Program under Grant No. JNFX2025067.}  	

\thanks{Muhammad Asif and Zhu Shoujin are with the School of Electrical and Information Engineering, Tongling University, Tongling 244002, China (e-mails: masif@tlu.edu.cn, 2023028@tlu.edu.cn).
	
 Asim Ihsan is with the Interdisciplinary Research Center for Communication Systems and Sensing, King Fahd University of Petroleum \& Minerals (KFUPM ) Dhahran, Saudi Arabia (e-mail: asim.ihsan@kfupm.edu.sa).
	
 Irfan Muhammad is with the Centre for Wireless Communications, University of Oulu, 90570 Oulu, Finland (e-mail: irfan.muhammad@oulu.fi).
  
M. H. N. Shaikh is with the Department of Electronic and Electrical Engineering, University College London, London WC1E 6BT, U.K (e-mail: hamza.commresearch@gmail.com).
	
S. T. Shah is with the School of Computer Science and Electronic Engineering, University of Essex, CO4 3SQ, Colchester, UK (e-mail: syed.shah@essex.ac.uk).
  	
 Symeon Chatzinotas is with the Interdisciplinary Centre for Security, Reliability and Trust (SnT), University of Luxembourg, 1855 Luxembourg City, Luxembourg (e-mail: symeon.chatzinotas@uni.lu).

}

\vspace{-0.7cm}}%

\markboth{}
{ \MakeLowercase{\textit{}}} 
\maketitle

\begin{abstract}
\leavevmode
To support joint information and energy delivery to energy-constrained users, this paper investigates a robust transmission design for a multi-user rate-splitting multiple access (RSMA)-based simultaneous wireless information and power transfer (SWIPT) system empowered by movable antennas (MAs) and a reconfigurable intelligent surface (RIS) in the presence of channel state information (CSI) uncertainty and residual hardware impairments (HIs). Unlike conventional fixed-position antenna systems, the effective channel responses in MAs-enabled systems are inherently dependent on antenna positions, causing CSI uncertainty to affect not only active and passive beamforming design but also antenna position adaptation. Furthermore, residual HIs arising from practical transceiver components alter the effective signal-to-interference-plus-noise ratios (SINRs) and introduce additional coupling among transmit beamforming, RIS reflection control, common-rate allocation, power-splitting ratio optimization, and antenna position adaptation. Thus, CSI uncertainty and HIs jointly influence all optimization variables of the system, giving rise to a highly coupled and significantly more challenging resource allocation problem. To address these challenges, a robust resource allocation framework is proposed to maximize the achievable sum-rate by jointly optimizing common-rate allocation, transmit beamforming, RIS reflection coefficients, power-splitting ratios, and MAs positions while satisfying practical system constraints. The resulting highly coupled non-convex optimization problem is addressed through a successive decomposition into active beamforming, RIS reflection design, power-splitting ratio optimization, and MAs position optimization subproblems, where tractable convex surrogate functions are constructed to obtain computationally efficient solutions. Additionally, to facilitate practical SWIPT operation, a non-linear energy harvesting model is incorporated into the system design. Simulation results demonstrate the effectiveness of the proposed framework and reveal significant performance gains over benchmark schemes in terms of achievable sum-rate, robustness, and convergence behavior.

\end{abstract}


\IEEEpeerreviewmaketitle


\section{Introduction}
\IEEEPARstart {T} {he} continuous expansion of internet-of-things (IoT) applications and the increasing number of interconnected smart devices have resulted in unprecedented growth in wireless traffic. Meeting the stringent performance requirements imposed by rising wireless traffic and massive device connectivity requires future wireless networks to deliver higher data rates, broader coverage, and reliable transmission with high energy efficiency and reduced hardware complexity \cite{wang2022gcwcn,sun2026cross}. Beyond these communication requirements, next-generation wireless networks are also expected to facilitate sustainable energy delivery to a large population of energy-constrained wireless devices. Motivated by this requirement, simultaneous wireless information and power transfer (SWIPT) has emerged as a promising technology that enables the concurrent transmission of information and wireless energy by exploiting radio-frequency (RF) signals \cite{zeng2017communications}. The ability of SWIPT to support both information transmission and energy delivery has attracted growing research interest across diverse wireless applications, including semantic communication \cite{lin2026semantic}, physical-layer security \cite{zhu2020robust}, and cognitive radio (CR) networks \cite{zhou2018artificial}.

However, conventional SWIPT systems rely on fixed-position antennas (FPA), limiting their ability to fully exploit the spatial characteristics of the wireless channel and consequently reducing the achievable diversity and multiplexing gains. To overcome this challenge, movable antennas (MAs) introduce a new transmission paradigm by enabling antenna elements to adapt their positions according to the wireless propagation environment \cite{zhu2023movable,xiao2024multiuser}. In MAs-assisted systems, each antenna element is flexibly connected to its corresponding RF chain, allowing it to relocate freely within a predefined movement region. This position flexibility introduces additional spatial degrees of freedom, enabling more favorable propagation conditions and thereby improving both diversity and spatial multiplexing gains \cite{ma2023mimo,zhu2023modeling}. Motivated by the advantages offered by MAs, several recent studies have investigated their integration into SWIPT networks \cite{huang2025weighted,ding2025weighted,chen2024movable,dong2025movable,gao2025movable}. For instance, the works in \cite{huang2025weighted, ding2025weighted} investigated MAs-assisted SWIPT networks and proposed weighted sum harvested power maximization frameworks, where the transmit precoding vectors and MAs positions are optimized simultaneously. The study presented in \cite{chen2024movable} explored the use of MAs in wireless-powered mobile-edge computing (MEC) systems, where the antenna deployment strategy was designed to maximize the overall computation rate. The work reported in \cite{dong2025movable} examined the application of MAs in SWIPT systems and demonstrated their effectiveness in strengthening physical-layer security. In \cite{gao2025movable}, MAs were incorporated into a wireless-powered non-orthogonal multiple access (NOMA)-based SWIPT system, where power allocation, time allocation, and MAs deployment were jointly designed to improve the achievable network throughput.
 
Besides MAs, rate-splitting multiple access (RSMA) has gained increasing attention for SWIPT systems due to its superior interference management capability, which facilitates reliable communication while improving spectral efficiency \cite{mao2022rate}. RSMA differs fundamentally from conventional space-division multiple access (SDMA) and NOMA by incorporating partial interference decoding, which enables dynamic allocation of transmit power to the common and private message streams \cite{clerckx2023primer,sun2025dual}. In particular, more power can be assigned to the common stream to improve energy harvesting performance, while the private streams ensure the satisfaction of individual QoS requirements. Consequently, RSMA offers greater flexibility in simultaneously supporting wireless energy transfer and information delivery, making it particularly suitable for SWIPT networks. Inspired by the advantages offered by RSMA, several recent works have investigated its incorporation into SWIPT systems \cite{karim2025finite, li2021full, garcia2024rate, galappaththige2024sum}. The authors in \cite{karim2025finite} considered an RSMA-assisted SWIPT network under finite block-length transmission and analyzed its communication performance. In \cite{li2021full}, cooperative RSMA was employed in a multicast SWIPT network, where the system transmit power was minimized through the joint optimization of active beamforming, message splitting, and power-splitting ratios. In \cite{garcia2024rate}, RSMA was incorporated into a cognitive radio SWIPT network to improve physical-layer security. In \cite{galappaththige2024sum}, the authors considered a cell-free massive multiple-input multiple-output (MIMO) SWIPT architecture employing RSMA and developed a transmission design to maximize the achievable system throughput.

Meanwhile, reconfigurable intelligent surface (RIS) facilitates programmable manipulation of wireless signal propagation through reconfigurable reflecting elements, thereby improving communication performance \cite{liu2021reconfigurable,asif2024securing}. Specifically, RIS achieves programmable control over wireless signal propagation by reconfiguring the reflection responses of its passive reflecting elements. This capability enables the reflected waves to be redirected toward the intended users, thereby enhancing channel quality and improving network coverage. \cite{ihsan2022energy, asif2026robust}. As a result, the nearly passive architecture of RIS enables wide-area coverage enhancement and improved network connectivity while maintaining low hardware cost and power consumption, making it an attractive option for SWIPT networks. Motivated by these advantages, several recent studies have investigated the integration of RIS into SWIPT networks \cite{yaswanth2023energy,ren2022ris,sharma2024robust,gao2023exploiting,zargari2021max}. Specifically, an RIS-enabled SWIPT framework was considered in \cite{yaswanth2023energy}, with the objective of minimizing transmit power while meeting the required QoS and harvested-energy levels. In \cite{ren2022ris}, RIS was integrated into a cooperative NOMA-based SWIPT framework, where the transmission strategy was developed to maximize the achievable communication rate. In \cite{sharma2024robust}, a robust energy efficiency maximization framework was developed for an RIS-assisted SWIPT network under channel estimation errors. The work in \cite{gao2023exploiting} considered a SWIPT system equipped with multiple RISs under interference channels and demonstrated the effectiveness of RIS deployment in improving the achievable sum-rate. In \cite{zargari2021max}, a max--min energy efficiency optimization problem was investigated for an RIS-assisted SWIPT system and demonstrated that RIS deployment can significantly enhance the minimum user energy efficiency.

Inspired by the aforementioned advantages of MAs, RSMA, and RIS, their integration into a unified framework has the potential to further improve SWIPT performance. By jointly leveraging antenna position flexibility, efficient interference management, and intelligent propagation control, the resulting system can achieve enhanced spectral efficiency, increased harvested energy, and more reliable service delivery. Consequently, the combined deployment of these technologies is expected to achieve performance gains beyond those attainable by each technique individually. In this context, recent studies have investigated the integration of MAs and RIS into SWIPT systems \cite{amiri2025movable} as well as the incorporation of RSMA and RIS into SWIPT networks \cite{zhang2023energy,asif2024leveraging,hashempour2024secure}. Specifically, \cite{amiri2025movable} considered an MAs-enabled simultaneously transmitting and reflecting RIS (STAR-RIS) SWIPT network based on an FPA architecture and addressed the joint maximization of achievable sum-rate and energy harvesting efficiency. In \cite{zhang2023energy}, an RIS-assisted RSMA-SWIPT network based on an FPA architecture was considered, where a proximal policy optimization (PPO)-based framework was developed to maximize the system energy efficiency. The study presented in \cite{asif2024leveraging} integrated STAR-RIS into an RSMA-SWIPT framework employing a FPA architecture, where the corresponding resource allocation strategy was developed to improve the achievable system throughput. In \cite{hashempour2024secure}, an optimization framework was developed to improve the physical-layer security of a STAR-RIS-assisted RSMA-SWIPT network based on an FPA architecture.

 In practical multiuser SWIPT deployments, propagation blockage and inter-user interference can limit the simultaneous delivery of information and energy. This challenge is particularly relevant to downlink networks serving energy-constrained terminals in obstructed indoor or outdoor environments, where walls, buildings, or other obstacles may hinder both communication and wireless energy delivery. In such settings, an RIS provides a controllable reflected path for both information transmission and energy delivery. Optimizing its reflection coefficients shapes the desired-signal and interference powers while influencing the RF power available for harvesting. MAs complement this capability by adapting the transmit antenna geometry and the BS--RIS channel responses. However, spatial adaptation alone does not necessarily eliminate inter-user interference, and suppressing interference solely for information decoding may reduce the power available for energy harvesting. RSMA provides additional flexibility by allowing users to decode a common stream before their private streams, while both common and private streams contribute to energy harvesting. The decoded common stream can therefore be cancelled in the information-decoding branch without removing its contribution to the energy-harvesting branch. These complementary roles motivate the joint design of antenna positions, RIS reflection coefficients, transmit precoders, common-rate allocation, and PS ratios to improve the achievable sum-rate while satisfying users' QoS and energy-harvesting requirements.

Although prior studies have demonstrated the effectiveness of advanced transmission technologies for improving the performance of SWIPT systems, the existing literature still suffers from the following limitations: \textbf{1)} Existing works have primarily focused on either MAs-assisted RIS-enabled SWIPT architectures \cite{amiri2025movable} or RIS-assisted RSMA-SWIPT frameworks \cite{zhang2023energy,asif2024leveraging,hashempour2024secure}. However, a comprehensive investigation of SWIPT systems jointly empowered by MAs, RSMA, and RIS is still lacking. In particular, the potential synergy arising from the joint exploitation of antenna mobility, flexible interference management, and intelligent propagation control for enhancing the performance of SWIPT networks remains largely unexplored. As a result, it remains unclear to what extent the complementary advantages of MAs, RSMA, and RIS can be jointly leveraged to enhance both wireless information transmission and energy transfer in SWIPT networks; \textbf{2)} Moreover, existing studies predominantly rely on idealized assumptions and overlook the joint impact of channel state information (CSI) uncertainty and transceiver hardware impairments (HIs). This limitation becomes particularly critical in the considered MAs-assisted RIS-enabled RSMA-SWIPT system, where jointly integrating MAs, RIS, and RSMA into the SWIPT framework under CSI uncertainty and residual HIs substantially increases the complexity of robust transmission design. Additionally, since antenna positions directly determine the effective channel responses, CSI uncertainty propagates not only through active beamforming and RIS reflection optimization, but also through antenna position adaptation. Consequently, CSI uncertainty may lead to inaccurate antenna position adaptation, thereby reducing the spatial adaptation gains. Furthermore, transceiver HIs distort both transmitted and received signals, alter the effective SINR expressions, and introduce additional coupling among beamforming design, RIS configuration, common-rate allocation, power-splitting ratio optimization, and antenna position adaptation. As a result, CSI uncertainty and HIs jointly affect all optimization variables of the system, giving rise to a highly coupled and significantly more challenging resource allocation problem. To the best of our knowledge, a robust transmission design for MAs-assisted RIS-enabled RSMA-SWIPT systems that jointly integrates MAs, RIS, and RSMA while enabling simultaneous wireless information transmission and energy transfer under CSI uncertainty and residual HIs has not been investigated in the existing literature.

The above research gaps motivate the present work, whose main contributions are summarized as follows.

\begin{itemize}
	\item We propose a robust transmission design for an RIS-empowered MAs-assisted multi-user RSMA-SWIPT system by explicitly accounting for CSI uncertainty and transceiver HIs. The considered framework captures the coupled impact of antenna-position-dependent channels, CSI uncertainty, and residual HIs, leading to a highly challenging joint optimization problem involving common-rate allocation, transmit beamforming, RIS reflection coefficients, power-splitting ratios, and MAs positions. Additionally, the practical behavior of the energy harvesting circuit is characterized through a non-linear energy harvesting model.
	
   \item We develop an alternating optimization algorithm that decomposes the joint design into transmit
   beamforming, RIS reflection, MAs position, and PS-ratio optimization subproblems. Tractable convex
   reformulations are constructed to solve the non-convex subproblems, while the PS and associated common-rate allocation subproblem admits a closed-form solution with an explicit feasibility condition. These updates jointly address the coupled design variables under CSI uncertainty and residual HIs.
   
   \item We provide analytical insights into the performance limits and the role of antenna-position optimization. Specifically, we characterize the high-power behavior of the common- and private-stream SINRs under fixed transmission configurations and establish a receiver-distortion-induced SINR upper bound. We also identify a single-transmit-path special case in which antenna-position-induced phase changes can be absorbed into the precoders under the stated assumptions, yielding equivalent optimal performance across feasible MAs layouts.
	
	\item Finally, extensive numerical results verify the effectiveness of the proposed robust design and highlight its superiority over existing benchmark schemes in terms of achievable sum-rate and robustness against CSI uncertainty and residual HIs. Moreover, the proposed algorithm exhibits fast convergence and stable performance under various system configurations.
\end{itemize}  
            
 \begin{figure}[!t]
	\centering
	\includegraphics [width=0.40\textwidth]{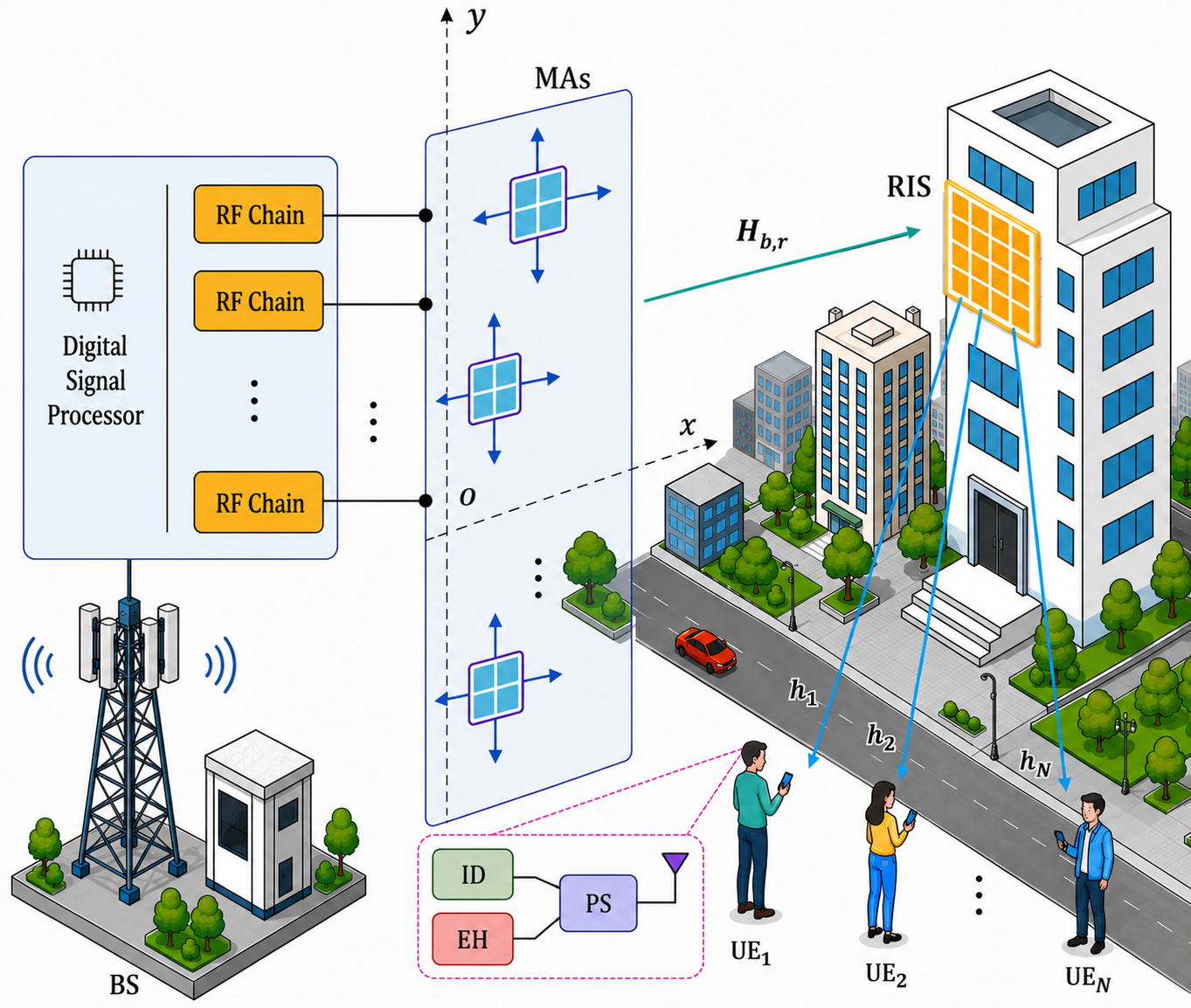}
	\caption{Illustration of system model.}
	\label{f1}
\end{figure} 
            
\noindent\textit{Notation:} Bold uppercase and lowercase Latin letters denote matrices and vectors, respectively. Greek symbols denoting matrices or vectors are also bold. Scalar quantities are nonbold, and sets are calligraphic.

\section{System Model and Problem Formulation}
We consider an RIS-assisted MAs-enabled multi-user RSMA-SWIPT network under CSI uncertainty and transceiver HIs, as depicted in Fig.~\ref{f1}. The BS utilizes $K$ MAs to simultaneously serve $N$ single-antenna users. Further, a power-splitting (PS) receiver is employed at each user to simultaneously perform information decoding (ID) and energy harvesting (EH) \cite{asif2026robust}. Additionally, due to severe blockage effects in the propagation environment, direct signal propagation from the BS to the users is assumed to be unavailable. Therefore, all user transmissions are established through an RIS consisting of $Q$ passive reflecting elements.

\subsection{Channel Model}
The wireless propagation environment is characterized using the planar far-field response model \cite{zhu2023modeling}. This model assumes that the spatial extent of the transmit antenna region is sufficiently small relative to the separation between the transmitter and the receiver. Accordingly, different propagation paths associated with the same channel share identical angular parameters and path gains, whereas the phase responses vary with the antenna positions.

Let
$\widehat{\boldsymbol{\zeta}}=\left[\widehat{\boldsymbol{\zeta}}_1,\widehat{\boldsymbol{\zeta}}_2,\ldots,\widehat{\boldsymbol{\zeta}}_K\right]\in\mathbb{R}^{2\times K}$
denotes the position matrix of the $K$ transmit MAs, where $\widehat{\boldsymbol{\zeta}}_k=[x_k,y_k]^T \in \mathcal Z_t$ represents the location of the $k$-th BS antenna for all $k\in\mathcal K=\{1,2,\ldots,K\}$. Here, $\mathcal Z_t$ denotes the feasible movement region available to the $K$ transmit MAs. Furthermore, let $K_t$ and $K_r$ denote the numbers of propagation paths observed at the transmit and receive sides, respectively. The elevation and azimuth angles corresponding to the $i$-th transmit-side propagation path are given by $\theta_i^{e}\in[0,\pi]$ and $\theta_i^{a}\in[0,\pi]$, respectively, where $i=1,2,\ldots,K_t$. Likewise, $\vartheta_j^{e}\in[0,\pi]$ and $\vartheta_j^{a}\in[0,\pi]$ denote the elevation and azimuth angles associated with the $j$-th receive-side propagation path, where $j=1,2,\ldots,K_r$. Then, the propagation distance difference between the reference origin and the antenna position $\widehat{\boldsymbol{\zeta}}_k$ for the $i$-th transmit path can be expressed as 
\begin{equation}
	\Xi\left(\widehat{\boldsymbol{\zeta}}_k,\theta_i^{{e}},\theta_i^{{a}}\right)
	=
	x_k \sin\theta_i^{{e}}\cos\theta_i^{{a}}
	+
	y_k \cos\theta_i^{{e}}.\label{1}
\end{equation}

Accordingly, for BS-RIS channel, the field response vector (FRV) associated with the BS antenna located at $\widehat{\boldsymbol{\zeta}}_k$ is given by
	\begin{align}
		\mathbf{v}^{\mathrm{t}}\left(\widehat{\boldsymbol{\zeta}}_k\right)
		=&\Big[e^{j\frac{2\pi}{\lambda}\Xi\left(\widehat{\boldsymbol{\zeta}}_k,\theta_1^e,\theta_1^a\right)},
		\ldots,
		e^{j\frac{2\pi}{\lambda}\Xi\left(\widehat{\boldsymbol{\zeta}}_k,\theta_{K_t}^{e},\theta_{K_t}^{a}\right)}\Big]^T
		\nonumber\\
		&\in\mathbb{C}^{K_t\times 1}.
		\label{2}
	\end{align}
where $\lambda$ is the carrier wavelength. Further, $\frac{2\pi}{\lambda}\Xi\left(\widehat{\boldsymbol{\zeta}}_k,\theta_i^{{e}},\theta_i^{{a}}\right)$, for $1\leq i\leq K_t$, corresponds to the phase shift introduced by the $i$-th propagation path with respect to the reference origin for the antenna located at $\widehat{\boldsymbol{\zeta}}_k$. Accordingly, the field response matrix (FRM) associated with the $K$ BS antennas is expressed as
\begin{equation}
	\mathbf{V}^{\mathrm{t}}(\widehat{\boldsymbol{\zeta}})
	\triangleq
	\left[
	\mathbf{v}^{\mathrm{t}}(\widehat{\boldsymbol{\zeta}}_1),
	\mathbf{v}^{\mathrm{t}}(\widehat{\boldsymbol{\zeta}}_2),
	\ldots,
	\mathbf{v}^{\mathrm{t}}(\widehat{\boldsymbol{\zeta}}_K)
	\right]
	\in
	\mathbb{C}^{K_t\times K}. \label{3}
\end{equation}

Similarly, the FRV corresponding to the $q$-th RIS reflecting element located at $\widehat{\boldsymbol{\tau}}_q=[x_q,y_q]^T,$ can be expressed as
	\begin{align}
		\mathbf{v}^{\mathrm{r}}\left(\widehat{\boldsymbol{\tau}}_q\right)
		=&\Big[e^{j\frac{2\pi}{\lambda}\Xi\left(\widehat{\boldsymbol{\tau}}_q,\vartheta_1^e,\vartheta_1^a\right)},
		\ldots, 
		e^{j\frac{2\pi}{\lambda}\Xi\left(\widehat{\boldsymbol{\tau}}_q,\vartheta_{K_r}^{e},\vartheta_{K_r}^{a}\right)}\Big]^T\nonumber\\
		&\in\mathbb{C}^{K_r\times 1}.\label{4}
	\end{align}
where $\frac{2\pi}{\lambda}\Xi\left(\widehat{\boldsymbol{\tau}}_q,\vartheta_j^e,\vartheta_j^a\right)$, for $j=1,2,\ldots,K_r$, denotes the phase variation of the $j$-th receive-side path induced by the RIS element position $\widehat{\boldsymbol{\tau}}_q$ relative to the RIS reference origin. Consequently, the corresponding RIS FRM is given by
\begin{equation}
	\mathbf{V}^{\mathrm{r}}(\widehat{\boldsymbol{\tau}})
	\triangleq
	\left[
	\mathbf{v}^{\mathrm{r}}(\widehat{\boldsymbol{\tau}}_1),
	\mathbf{v}^{\mathrm{r}}(\widehat{\boldsymbol{\tau}}_2),
	\ldots,
	\mathbf{v}^{\mathrm{r}}(\widehat{\boldsymbol{\tau}}_Q)
	\right]
	\in
	\mathbb{C}^{K_r\times Q}.\label{5}
\end{equation}

Next, let $\boldsymbol{\Omega}\in\mathbb{C}^{K_r\times K_t}$
denote the path response matrix associated with the BS--RIS link. Then, the equivalent BS--RIS channel, denoted by $\mathbf{H}_{b,r}$, is given by
\begin{equation}
	\mathbf{H}_{b,r}(\widehat{\boldsymbol{\zeta}})
	=
	\mathbf{V}^{\mathrm{r}}(\widehat{\boldsymbol{\tau}})^H
	\boldsymbol{\Omega}
	\mathbf{V}^{\mathrm{t}}(\widehat{\boldsymbol{\zeta}})
	\in
	\mathbb{C}^{Q\times K}.	\label{6}
\end{equation}

Let $\mathbf{h}_{r,n}\in\mathbb{C}^{Q\times 1}$ denote the channel vector associated with the RIS-to-user link of the $n$-th user. The corresponding estimated BS--RIS--user cascaded
channel is expressed as
\begin{equation}
	\mathbf{f}_{n}^H(\widehat{\boldsymbol{\zeta}})
	=
	\mathbf{h}_{r,n}^H
	\boldsymbol{\Theta}
	\mathbf{H}_{b,r}(\widehat{\boldsymbol{\zeta}})
	\in
	\mathbb{C}^{1\times K},\label{7}
\end{equation}
where $\boldsymbol{\Theta}=\operatorname{diag}\left\{[e^{j\varphi_1},e^{j\varphi_2},\ldots,e^{j\varphi_Q}]^T\right\}$ represents the RIS reflection coefficient matrix, with $\varphi_i \in [0,2\pi]$, $\forall i \in \mathcal{Q}=\{1,2,\ldots,Q\}$, denoting the adjustable phase shift of the $i$-th RIS element.

 \noindent{\textbf{Remark 1} (Extension to direct links):} \textit{The blocked direct-link assumption represents a deployment scenario with severe BS--user blockage. If a residual direct link exists, the effective channel becomes $\mathbf f_{n,\mathrm{tot}}^H(\widehat{\boldsymbol{\zeta}}) =\mathbf h_{b,n}^H(\widehat{\boldsymbol{\zeta}}) +\mathbf h_{r,n}^H\boldsymbol{\Theta} \mathbf H_{b,r}(\widehat{\boldsymbol{\zeta}})$, where $\mathbf h_{b,n}(\widehat{\boldsymbol{\zeta}})$ denotes
the position-dependent direct BS--user channel. The effective-channel uncertainty model can then be applied to the combined channel, with an appropriately specified uncertainty bound, and the SINR, hardware-distortion, and harvested-energy expressions are updated accordingly. The active beamforming and PS-ratio subproblems can retain their overall structure, with the rate and harvested-energy expressions evaluated using the combined effective channel. The RIS optimization subproblem would be updated to account for the combined direct and reflected signals, while the gradients with respect to the MAs positions would also include the dependence of the direct channel on antenna positions. These modifications preserve the overall alternating optimization structure and the main contribution of jointly designing robust beamforming, RIS reflection coefficients, PS ratios, and MAs positions.}

\subsection{Signal Model}
Following the RSMA transmission framework, the information intended for each user is partitioned into a common part and a private part prior to transmission. In particular, the private parts associated with the $N$ users are independently encoded into $N$ private data streams, represented by $\{\hat{p}_n\}_{n\in\mathcal N}$, where $\mathcal N=\{1,2,\ldots,N\}$. Meanwhile, the common parts intended to be decoded by all users are jointly combined and encoded into a common data stream, represented by $\hat{p}_c$. Accordingly, the transmitted signal generated at the BS can be expressed as
\begin{equation}
	\mathbf{x}
	=
	\mathbf{b}_c \hat{p}_c
	+
	\sum_{n=1}^{N} \mathbf{b}_n \hat{p}_n + \boldsymbol{\Upsilon}_t,\label{8}
\end{equation}
where $\mathbf{b}_n \in \mathbb{C}^{K\times 1}$ and $\mathbf{b}_c \in \mathbb{C}^{K\times 1}$ represent the beamforming vectors assigned to the private and common streams, respectively. All transmitted data streams are considered mutually independent and have unit average power, satisfying $\mathbb{E}\left[|\hat{p}_{n}|^2\right]=1$, $\forall n\in\mathcal N$, and $\mathbb{E}\left[|\hat{p}_{c}|^2\right]=1$. Additionally, the residual HI at the transmitter are characterized by the additive distortion noise vector $\boldsymbol{\Upsilon}_t$, which is modeled as
\begin{equation}
	\boldsymbol{\Upsilon}_t
	\sim
	\mathcal{CN}
	\!\Bigg(
	\mathbf{0},
	\varrho_t
	\widetilde{\mathrm{diag}}
	\!\Bigg(
	\sum_{n=1}^{N}\mathbf{b}_{n}\mathbf{b}_{n}^{H}
	+
	\mathbf{b}_{c}\mathbf{b}_{c}^{H}
	\Bigg)
	\Bigg),\label{9}
\end{equation}
following the widely adopted residual HIs model \cite{bjornson2014massive}. Specifically, $\boldsymbol{\Upsilon}_t$ is characterized as a zero-mean circularly symmetric complex Gaussian random vector, whose covariance matrix scales proportionally with the intended transmit signal power. Here, $\widetilde{\operatorname{diag}}(\cdot)$ denotes the operator that retains the diagonal entries of its matrix argument and sets all off-diagonal entries to zero. Further, $\varrho_t\in(0,1)$ characterizes the severity of the transmitter hardware impairments and corresponds to the ratio between the distortion noise power and the intended transmit signal power \cite{bjornson2014massive,shen2020beamforming}.

Accordingly, the signal received at user $n$ before power splitting is given by
\begin{align}
	\widetilde y_n
	=
	\overline{\mathbf f}_n^H
	(\widehat{\boldsymbol\zeta})\mathbf x+\nu_n,
	\quad \forall n\in\mathcal N,
	\label{10}
\end{align}
where $\overline{\mathbf f}_n$ denotes the actual effective channel and $\nu_n\sim\mathcal{CN}(0,\bar{\sigma}_n^2)$ is the receiver noise at user $n$. At each user, the received signal is divided between the ID and EH circuits through a PS receiver. The PS ratio $\eta_n\in[0,1]$ specifies the portion of received power used for ID, with the remaining portion
$1-\eta_n$ supplied to the EH circuit.

The ID branch experiences residual receiver distortion after power splitting, which is modeled as \cite{asif2024leveraging,bjornson2014massive}
\begin{equation}
	\Upsilon_{r,n}
	\sim
	\mathcal{CN}\!\left(
	0,\varrho_r\eta_n
	\mathbb E[|\widetilde y_n|^2]
	\right),
	\quad \forall n\in\mathcal N, \label{11}
\end{equation}
where $\varrho_r\in(0,1)$ characterizes the severity of the receiver hardware impairments\cite{shen2020beamforming}. It should be noted that imperfect hardware components introduce impairments such as phase noise, amplifier nonlinearities, I/Q imbalance, quantization errors, and DC offsets. In this work, the combined residual effects of transmitter and receiver HIs are modeled by the additive Gaussian distortion terms $\boldsymbol{\Upsilon}_t$ and $\Upsilon_{r,n}$ in \eqref{9} and \eqref{11}, respectively \cite{bjornson2014massive}. The transmitter distortion power scales with the intended signal power at each BS antenna through $\varrho_t$, while the receiver distortion power scales with $\eta_n\mathbb E[|\widetilde y_n|^2]$ through $\varrho_r$. These parameters characterize the combined residual distortion at each side, without modeling individual impairment mechanisms separately \cite{asif2024leveraging, bjornson2014massive}.

Accordingly, the signals delivered to the ID and EH circuits are respectively written as
\begin{align}
	y_n^{\rm ID}
	&=
	\sqrt{\eta_n}\,\widetilde y_n
	+\Upsilon_{r,n}
	+\widetilde n_d,
	\label{12}\\
	y_n^{\rm EH}
	&=
	\sqrt{1-\eta_n}\,\widetilde y_n.
	\label{13}
\end{align}
where $\tilde{n}_{\mathrm{d}} \sim \mathcal{CN}\big(0, \tilde{\sigma}^2_{\mathrm{d}}\big)$ denotes the additive Gaussian noise at the ID circuit.

Subsequently, channel estimation uncertainty is incorporated by expressing the actual effective channel of the $n$-th user as $\overline{\mathbf f}_n=\mathbf f_n+\Delta\mathbf f_n$ \cite{li2022robust,zhang2023robust}, where $\mathbf f_n$ denotes the estimated channel available
at the transmitter and $\Delta\mathbf f_n$ is the channel estimation error. Further, define
$\overline{\mathbf F}_n =\overline{\mathbf f}_n\overline{\mathbf f}_n^H$ and $\mathbf F_n=\mathbf f_n\mathbf f_n^H$ as the actual and estimated channel outer-product matrices, respectively. Accordingly, the actual channel outer-product matrix is expressed as $\overline{\mathbf F}_n=\mathbf F_n+\Delta\mathbf F_n$, where $\Delta\mathbf F_n$ denotes the outer-product estimation error matrix. We adopt the bounded matrix-error model $\|\Delta\mathbf F_n\|_2\leq\mu_n^2$, with $\mu_n^2=\mu\|\mathbf F_n\|_2$ \cite{asif2026robust,zheng2023zero}, where $\mu\in[0,1)$ controls the relative uncertainty level. Although $\mu$ is fixed, $\mu_n^2$ depends on the BS antenna positions and RIS coefficients through the estimated cascaded channel matrix.

To account for imperfect successive interference cancellation (SIC) \cite{singh2023rsma}, a fraction $\delta_n\in[0,1]$ of the received common-stream power remains as interference during private-stream decoding. The common- and private-stream SINRs, $\overline{\Gamma}_n^c$ and $\overline{\Gamma}_n^p$, are given in~\eqref{14} and~\eqref{15}, respectively, where $\sigma_n^2=\eta_n(1+\varrho_r)\bar{\sigma}_n^2 +\widetilde{\sigma}_{\mathrm d}^2$. Then, the achievable rates for the common and private streams decoded by the $n$-th user are respectively given by
\begin{figure*} [!t]
	\begin{align} \label{14}
		\overline{\Gamma}^{c}_{n} 
		= \frac{ \eta_n \operatorname{tr}\!\Big(\mathbf{b}_c^H\big({\mathbf{F}}_{n} + \Delta {\mathbf{F}}_{n}\big) \mathbf{b}_c\Big) }
		{\eta_n \operatorname{tr}\!\Bigg(
			\Bigg(
			(1+\varrho_r)\sum_{j=1}^N \mathbf{b}_j \mathbf{b}_j^{\!H}
			\;+\; \varrho_r\, \mathbf{b}_c \mathbf{b}_c^{\!H}
			\;+\; (1+\varrho_r)\varrho_t\, \widetilde{\operatorname{diag}}\!\Big(\sum_{j=1}^N \mathbf{b}_j \mathbf{b}_j^{\!H} + \mathbf{b}_c \mathbf{b}_c^{\!H}\Big)
			\Bigg) \Big({\mathbf{F}}_{n} + \Delta {\mathbf{F}}_{n}\Big)
			\Bigg)
			+ {\sigma}_n^2 },
	\end{align}
\end{figure*}

\begin{figure*} [!t] 
	\begin{align} \label{15}
		\overline{\Gamma}^{p}_{n} 
		= \frac{ \eta_n \operatorname{tr}\!\Big(\mathbf{b}_n^H\big({\mathbf{F}}_{n} + \Delta {\mathbf{F}}_{n}\big) \mathbf{b}_n\Big) }
		{\eta_n \operatorname{tr}\!\Bigg(
			\Bigg(
			\sum_{i \neq n} \mathbf{b}_i \mathbf{b}_i^{\!H}
			+\varrho_r\sum_{j=1}^N \mathbf{b}_j \mathbf{b}_j^{\!H}
			\;+\; (\varrho_r+\delta_n)\mathbf b_c\mathbf b_c^H
			\;+\; (1+\varrho_r)\varrho_t\, \widetilde{\operatorname{diag}}\!\Big(\sum_{j=1}^N \mathbf{b}_j \mathbf{b}_j^{\!H} + \mathbf{b}_c \mathbf{b}_c^{\!H}\Big)
			\Bigg) \Big({\mathbf{F}}_{n} + \Delta {\mathbf{F}}_{n}\Big)
			\Bigg)
			+ {\sigma}_n^2 },
	\end{align}
\end{figure*}

\begin{equation}
	R_n^c=\log_2\left(1+\overline{\Gamma}_n^c\right),
	\quad \forall n\in\mathcal N,
	\label{16}
\end{equation}
\begin{equation}
	R_n^p=\log_2\left(1+\overline{\Gamma}_n^p\right),
	\quad \forall n\in\mathcal N.
	\label{17}
\end{equation}

Next, the achievable sum-rate of the considered system can be expressed as
	\begin{align}
	{R}_{t}= & \sum\limits_{{\substack{n \in \mathcal N}}} \Big({r}^c_n + {R}^p_{n}\Big),   \label{18}
\end{align}
where ${r}^c_n$ denotes the allocated common rate for the $n$-th user.

Next, to accurately characterize the practical behavior of EH circuits, we employ a non-linear energy harvesting (EH) model that more effectively captures the saturation characteristics of EH hardware. The adopted model is expressed as follows \cite{xiong2017rate}:
\begin{equation}
		\overline{\varepsilon}_{n} =
	\frac{\widehat{\Omega}(\widehat{\varepsilon}_{n})-\widehat{W}_n\widehat{Y}_n}{1-\widehat{Y}_n},
	\quad \forall n\in\mathcal N,
	\label{20}
\end{equation}
\begin{figure*}[!t]
\begin{align}
	\widehat{\varepsilon}_n
	={}&
	(1-\eta_n)
	\operatorname{tr}\!\Bigg[
	\Bigg(
	\sum_{j=1}^{N}\mathbf b_j\mathbf b_j^H
	+\mathbf b_c\mathbf b_c^H+
	\varrho_t\widetilde{\operatorname{diag}}\!\left(
	\sum_{j=1}^{N}\mathbf b_j\mathbf b_j^H
	+\mathbf b_c\mathbf b_c^H
	\right)
	\Bigg)
	(\mathbf F_n+\Delta\mathbf F_n)
	\Bigg]
	+\hat{\sigma}_n^2,
	\label{18b}
\end{align}
	\hrulefill
\end{figure*}
where $\widehat{Y}_n=\frac{1}{1+e^{\widehat{w}_n \widehat{y}_n}}$ and $\widehat{\Omega}(\widehat{\varepsilon}_{n})=\frac{\widehat{W}_n}{1+e^{-\widehat{w}_n(\widehat{\varepsilon}_{n}-\widehat{y}_n)}}$. Here, $\widehat{W}_n$ denotes the maximum harvested power of the EH circuit, while $\widehat{\varepsilon}_{n}$ is given in \eqref{18b}, with $\hat{\sigma}_n^2=(1-\eta_n)\bar{\sigma}_n^2$. Moreover, $\widehat{w}_n$ and $\widehat{y}_n$ are circuit-dependent parameters that characterize the non-linear behavior of the EH circuit \cite{asif2026robust,xiong2017rate}.

\subsection{Problem Formulation}
The considered optimization framework seeks to maximize the achievable system sum-rate in the presence of transceiver HIs and CSI uncertainty by jointly designing the common and private precoders $\mathbf{b}_c$ and $\mathbf{b}_n$, $\forall n\in\mathcal N$, together with the RIS scattering matrix $\boldsymbol{\Theta}$, PS ratios $\{\eta_n\}_{n\in\mathcal N}$, and the MAs positions $\widehat{\boldsymbol{\zeta}}$. Accordingly, the corresponding optimization problem can be formulated as
\begin{subequations}\label{P1}
	\begin{align}
		\text{\textbf{(P1)}}	& \max_{\substack{
				\mathbf b_c,\widehat{\boldsymbol{\zeta}},\boldsymbol{\Theta},\{\mathbf b_n,r_n^c,\eta_n\}_{n\in\mathcal N}
		}} \    \sum\limits_{{\substack{n=1}}}^N \Big({r}^c_n + {R}^p_{n}\Big),  \label{21a}  \\
		s.t.\ C_1:\ & {r}^c_n + {R}^p_{n}  \geq  {R}_{\min}, \ \forall n\in \mathcal N, \label{21b} 
		\end{align}
		\begin{align}
		\ C_2:\ &  \sum\limits_{{\substack{n=1}}}^N  {r}^c_n  \leq {R}^c_{n}, \ \forall n\in \mathcal N, \label{21c}\\
		\ C_3:\ & (1+\varrho_t)\bigg(\sum_{n=1}^{N}\|\mathbf b_n\|_2^2+\|\mathbf b_c\|_2^2\bigg)\leq P_{\max}, \label{21d}\\
		\ C_4:\ & \widehat{\kappa}_n (1-\widehat{Y}_n) \leq \widehat{\Omega}(\widehat{\varepsilon}_{n})-\widehat{W}_n\widehat{Y}_n, \forall n\in \mathcal N \label{21e}\\
		\ C_5:\ & \left|e^{j\varphi_i} \right|  = 1, \ \forall i\in \mathcal Q, \label{21f} \\
		\ C_6: \ &\varphi_i \in [0, 2\pi],  \ \forall i\in \mathcal Q, \label{21g}\\
		\ C_7: \ &\left\|\widehat{\boldsymbol{\zeta}}_k - \widehat{\boldsymbol{\zeta}}_j\right\|_2 \geq D_{\min}, \ \forall k,j\in \mathcal K, \quad k \neq j, \label{21h}\\
		\ C_8: \ & \widehat{\boldsymbol{\zeta}} \in \mathcal{Z}_t, \label{21i}\\
		\ C_9: \ &0 \leq\eta_n \leq 1, \forall n\in \mathcal N, \label{21j}\\
		\ C_{10}:  \ & r_n^c\geq0,\quad \forall n\in \mathcal N. \label{21k}
	\end{align}
\end{subequations}
where $D_{\min}$ denotes the minimum inter-antenna separation imposed to limit the impact of mutual coupling, and $\widehat{\kappa}_n$ denotes the minimum harvested energy requirement. Moreover, $P_{\max}$ and $R_{\min}$ represent the total radiated-power budget at the BS and the minimum QoS rate requirement, respectively. Constraint $C_1$ guarantees the QoS requirements of the users, while constraint $C_2$ ensures reliable decoding of the common stream. Constraints $C_4$ and $C_5$ enforce the EH requirement and the unit-modulus constraint on the RIS reflecting elements, respectively. Furthermore, constraint $C_7$ preserves adequate spacing among MAs to avoid mutual coupling, whereas constraint $C_8$ restricts the MA positions within the feasible region.

\section{Proposed Optimization Framework}
The intricate coupling among optimization variables ${r}_n^{c}$, $\eta_n$, $\mathbf{b}_{c}$, $\mathbf{b}_{n}$, $\boldsymbol{\Theta}$, and $\widehat{\boldsymbol{\zeta}}$, gives rise to a highly non-convex optimization problem, making \textbf{(P1)} computationally challenging to handle. Thus, to facilitate the subsequent optimization process, problem \textbf{(P1)} is transformed into a conservative robust formulation. The following theorem provides the foundation for the subsequent reformulation:

\noindent \textbf{Theorem 1:} 
Suppose that $\widehat{\mathbf{Y}}$ and $\widehat{\mathbf{T}}$ are Hermitian matrices, and that $\widehat{\mathbf{Y}}$ is a rank-one positive semi-definite (PSD) matrix. Under the norm-bounded constraint $\|\widehat{\mathbf{T}}\| \leq \mu^2$, the following relation holds:
\begin{equation}
	\max_{\|\widehat{\mathbf{T}}\| \leq \mu^2} 
	\operatorname{tr}(\widehat{\mathbf{Y}}\widehat{\mathbf{T}})
	=
	\mu^2 \operatorname{tr}(\widehat{\mathbf{Y}}).
	\label{22}
\end{equation}

\noindent \textbf{Proof:} 
Consider two arbitrary complex matrices $\widehat{\mathbf{R}}$ and $\widehat{\mathbf{P}}$. By exploiting the definition of the dual norm, the maximum inner product generated by $\widehat{\mathbf{R}}$ over the bounded set $\|\widehat{\mathbf{R}}\|\leq 1$ is characterized by the dual norm of $\widehat{\mathbf{P}}$, which yields
\begin{equation}
	\max_{\|\widehat{\mathbf{R}}\| \leq 1}
	\langle \widehat{\mathbf{R}}, \widehat{\mathbf{P}} \rangle
	=
	\|\widehat{\mathbf{P}}\|_{\mathrm{d}}.
	\label{23}
\end{equation}

According to Hölder's inequality, the trace operation satisfies
\begin{equation}
	\operatorname{tr}(\widehat{\mathbf{P}}^{H}\widehat{\mathbf{R}})
	\leq
	\|\widehat{\mathbf{R}}\|
	\|\widehat{\mathbf{P}}\|_{\mathrm{d}},
	\label{24}
\end{equation}
where $\|\cdot\|_{\mathrm{d}}$ denotes the corresponding dual norm.

Next, let $\widehat{\mathbf{Y}}$ and $\widehat{\mathbf{T}}$ be Hermitian matrices satisfying the bounded condition $\|\widehat{\mathbf{T}}\|\leq \mu^{2}$. Then, from \eqref{24}, it follows that
\begin{equation}
	\operatorname{tr}(\widehat{\mathbf{Y}}\widehat{\mathbf{T}})
	\leq
	\|\widehat{\mathbf{T}}\|
	\|\widehat{\mathbf{Y}}\|_{\mathrm{d}}
	\leq
	\mu^{2}\|\widehat{\mathbf{Y}}\|_{\mathrm{d}}.
	\label{25}
\end{equation}

Hence, the following expression can be directly obtained:
\begin{equation}
	\max_{\|\widehat{\mathbf{T}}\| \leq \mu^{2}}
	\operatorname{tr}(\widehat{\mathbf{Y}}\widehat{\mathbf{T}})
	=
	\mu^{2}\|\widehat{\mathbf{Y}}\|_{\mathrm{d}}.
	\label{26}
\end{equation}

Since the spectral norm and nuclear norm form a dual pair, the dual norm of $\widehat{\mathbf{Y}}$ can be represented by its nuclear norm, i.e.,
\begin{equation}
	\|\widehat{\mathbf{Y}}\|_{\mathrm{d}}
	=
	\|\widehat{\mathbf{Y}}\|_{\mathrm{n}}
	=
	\sum_{m}\beta_{m},
	\label{27}
\end{equation}
where $\beta_{m}$ denotes the eigenvalues of $\widehat{\mathbf{Y}}$, and $\|\cdot\|_{\mathrm{n}}$ represents the nuclear norm.

Further, by exploiting the rank-one PSD property of $\widehat{\mathbf{Y}}$, it follows that its nuclear norm is equal to its trace. Accordingly,
\begin{equation}
	\|\widehat{\mathbf{Y}}\|_{\mathrm{d}} = \|\widehat{\mathbf{Y}}\|_{\mathrm{n}}
	=
	\operatorname{tr}(\widehat{\mathbf{Y}}).\label{28}
\end{equation}

Substituting the above relation into \eqref{26} yields
\begin{equation}
	\max_{\|\widehat{\mathbf{T}}\| \leq \mu^{2}}
	\operatorname{tr}(\widehat{\mathbf{Y}}\widehat{\mathbf{T}})
	=
	\mu^{2}\operatorname{tr}(\widehat{\mathbf{Y}}).
	\label{29}
\end{equation}

This completes the proof.

Next, by defining $\mathbf{B}_n=\mathbf{b}_n\mathbf{b}_n^{H}$ and
$\mathbf{B}_c=\mathbf{b}_c\mathbf{b}_c^{H}$ as rank-one PSD matrices, \textbf{Theorem~1} can be applied to obtain
\begin{equation}
	\max _{\|\Delta {\mathbf{F}}_{n}\| \leq {\mu}_n^2} \operatorname{tr}(\Delta {\mathbf{F}}_{n} \mathbf{B}_c)={\mu}_n^2 \operatorname{tr}(\mathbf{B}_c), \label{30}
\end{equation} 
and,
\begin{equation}
	\max _{\|\Delta {\mathbf{F}}_{n}\| \leq {\mu}_n^2} \operatorname{tr}(\Delta {\mathbf{F}}_{n} \mathbf{B}_n)={\mu}_n^2 \operatorname{tr}(\mathbf{B}_n). \label{31}
\end{equation} 

Accordingly, conservative lower bounds on the common- and private-stream rates in \eqref{16} and \eqref{17}, respectively, are given by
\begin{equation}
	\bar{R}^c_{n}=\log _2\Bigg(1+\frac{\eta_n\operatorname{tr}\Big(\Big({\mathbf{F}}_{n}-\mu_n^2 \mathbf{I}\Big)\mathbf{B}_c \Big) }
	{\eta_n\operatorname{tr}\!\Big(
		\Big({\mathbf{F}}_{n}+\mu_n^2 \mathbf{I}\Big) \boldsymbol{\Psi}_1
		\Big)
		+ {\sigma}_n^2 }\Bigg), \label{32}
\end{equation}

\begin{equation}
\bar{R}^p_{n}=\log _2\Bigg(1+\frac{\eta_n\operatorname{tr}\Big(\Big({\mathbf{F}}_{n} - \mu_{n}^2 \mathbf{I}\Big)\mathbf{B}_n \Big) }
{\eta_n\operatorname{tr}\!\Big(
	\Big({\mathbf{F}}_{n}+\mu_{n}^2 \mathbf{I}\Big) \boldsymbol{\Psi}_2
	\Big)
	+ {\sigma}_n^2 }\Bigg), \label{33}
\end{equation}
where $\boldsymbol{\Psi}_1
=(1+\varrho_r)\sum_{j=1}^{N}\mathbf B_j
+\varrho_r\mathbf B_c
+(1+\varrho_r)\varrho_t
\widetilde{\operatorname{diag}}\!\left(
\sum_{j=1}^{N}\mathbf B_j+\mathbf B_c\right)$
and $\boldsymbol{\Psi}_2
=\sum_{i\neq n}\mathbf B_i
+\varrho_r\sum_{j=1}^{N}\mathbf B_j
+(\varrho_r+\delta_n)\mathbf B_c
+(1+\varrho_r)\varrho_t
\widetilde{\operatorname{diag}}\!\left(
\sum_{j=1}^{N}\mathbf B_j+\mathbf B_c\right)$.

Further, a sufficient condition for satisfying the EH constraint in \eqref{21e} under the bounded matrix
uncertainty is given by
\begin{align}
	\big(1-\eta_n\big)
	\operatorname{tr}\Big(\!\big(\mathbf{F}_{n}
	-\mu_n^2\mathbf{I}\big)
	\boldsymbol{\Psi}_{\rm EH}\Big)
	+\hat{\sigma}_n^2
	\geq A_n,
	\label{34}
\end{align}
where
$\boldsymbol{\Psi}_{\rm EH}\triangleq
\sum_{j=1}^{N}\mathbf B_j+\mathbf B_c
+\varrho_t\widetilde{\operatorname{diag}}\!\left(
\sum_{j=1}^{N}\mathbf B_j+\mathbf B_c
\right)$, and
\begin{align}
	A_n = \widehat{y}_n
	-\frac{1}{\widehat{w}_n}\ln \Bigg(
	\frac{\widehat{W}_n}
	{\widehat{W}_n \widehat{Y}_n
		+(1-\widehat{Y}_n)\widehat{\kappa}_n}
	-1 \Bigg).
	\label{35}
\end{align}

Consequently, a robust reformulation of problem \textbf{(P1)} is given by

\begin{subequations}\label{P2}
	\begin{align}
		\text{\textbf{(P2)}}	& \max_{\substack{
				\mathbf B_c,\widehat{\boldsymbol{\zeta}},\boldsymbol{\Theta},\{\mathbf B_n,r_n^c,\eta_n\}_{n\in\mathcal N}
		}} \    \sum\limits_{{\substack{n=1}}}^N \Big({r}^c_n + \bar{R}^p_{n}\Big),  \label{36a}  \\
        s.t.\ \ &  {R}_{\min} \leq {r}^c_n + \bar{R}^p_{n}, \ \forall n\in \mathcal N, \label{36b} \\
       \ \ &  \sum\limits_{{\substack{n=1}}}^N  {r}^c_n  \leq \bar{R}^c_{n}, \ \forall n\in \mathcal N, \label{36c}\\
		\ \ & (1+\varrho_t)\left(\sum_{n=1}^{N}\operatorname{tr}(\mathbf B_n)+\operatorname{tr}(\mathbf B_c)\right)\leq P_{\max}, \label{36d}\\
		\ \ & \mathbf{B}_c \succeq 0, \ \mathbf{B}_n \succeq 0, \ \forall n\in \mathcal N, \label{36e}\\ 
		\ \ & \operatorname{rank}(\mathbf{B}_c)=1, \ \operatorname{rank}(\mathbf{B}_n)=1,  \forall n\in \mathcal N. \label{36f}\\
		\ \ & \operatorname{tr}\!\left[(\mathbf F_n-\mu_n^2\mathbf I)\mathbf B_c \right]\geq 0,
		\qquad \forall n\in\mathcal N, \label{36g}\\
		\ \ & \operatorname{tr}\!\left[(\mathbf F_n-\mu_n^2\mathbf I)\mathbf B_n\right]\geq 0,
		\qquad \forall n\in\mathcal N, \label{36h}\\
		\  \ & \eqref{21f}-\eqref{21k}, \eqref{34}. \label{36i} 
	\end{align}
\end{subequations}

\subsection{Updating Precoding Vectors}
Given the RIS configuration $\boldsymbol{\Theta}$, PS ratios $\{\eta_n\}_{n\in\mathcal N}$, and MAs positions $\widehat{\boldsymbol{\zeta}}$, the joint optimization problem reduces to the following transmit precoding design subproblem:
\begin{subequations}\label{P3}
	\begin{align}
		\text{\textbf{(P3)}}	& \max_{\mathbf B_c,\{\mathbf B_n,r_n^c\}_{n\in\mathcal N}} \    \sum\limits_{{\substack{n=1}}}^N \Big({r}^c_n + \bar{R}^p_{n}\Big),  \label{37a}   \\
		s.t.\ \ &  \eqref{21k}, \eqref{34},\eqref{36b} - \eqref{36h}. \label{37b} 
	\end{align}
\end{subequations}

The non-convex objective function and constraints render problem \textbf{(P3)} challenging to solve directly. To facilitate the subsequent convex reformulation, an auxiliary slack variable $\tau_n^p$ is introduced as
\begin{subequations}
\begin{align}
	& {r}^c_n + \tau_n^p \geq {R}_{\min}, \ \forall n\in \mathcal N,  \label{38a}
\end{align}
\begin{align}
	&\log_{2}\Big(\eta_n \operatorname{tr}\Big(\Big({\mathbf{F}}_{n} - \mu_{n}^2 \mathbf{I}\Big)\mathbf{B}_n \Big)+ \eta_n\operatorname{tr}\!\Big(
	\Big({\mathbf{F}}_{n}+\mu_{n}^2 \mathbf{I}\Big) \boldsymbol{\Psi}_2
	\Big)+\nonumber\\
	& {\sigma}_n^2 \Big)-\log_{2}\left(\eta_n\operatorname{tr}\!\Big(
	\Big({\mathbf{F}}_{n}+\mu_{n}^2 \mathbf{I}\Big) \boldsymbol{\Psi}_2
	\Big)+ {\sigma}_n^2  \right) \geq \tau_n^p.  \label{38b}
\end{align}
\end{subequations}

The non-convexity of \eqref{38b} arises from the difference between two concave functions. Thus, we define an auxiliary variable $\alpha^p_n$ as follows
\begin{align}
	&\log_{2}\Big(\eta_n \operatorname{tr}\Big(\Big({\mathbf{F}}_{n} - \mu_{n}^2 \mathbf{I}\Big)\mathbf{B}_n \Big)+ \eta_n \operatorname{tr}\!\Big(
	\Big({\mathbf{F}}_{n}+\mu_{n}^2 \mathbf{I}\Big) \boldsymbol{\Psi}_2
	\Big)+\nonumber\\
	 &{\sigma}_n^2 \Big)- \alpha^p_n \geq \tau_n^p , \forall n\in \mathcal N,  \label{39}
\end{align}
\begin{align}
	&\log_{2}\Big(\eta_n\operatorname{tr}\!\Big(
	\Big({\mathbf{F}}_{n}+\mu_{n}^2 \mathbf{I}\Big) \boldsymbol{\Psi}_2
	\Big)+ {\sigma}_n^2  \Big) \leq  \alpha^p_n.  \label{40}
\end{align}

Further, to facilitate the convex reformulation of \eqref{40}, an auxiliary variable $\varsigma_n$ is introduced as
\begin{align}
	&\log_{2}\left(\varsigma_n \right) \leq  \alpha^p_n, \forall n\in \mathcal N, \label{41}
\end{align}
\begin{align}
	&\eta_n\operatorname{tr}\!\Big(
	\Big({\mathbf{F}}_{n}+\mu_{n}^2 \mathbf{I}\Big) \boldsymbol{\Psi}_2
	\Big)+ {\sigma}_n^2 \leq  \varsigma_n, \forall n\in \mathcal N .  \label{42}
\end{align}

To facilitate the convexification of \eqref{41}, we employ the following first-order upper-bound approximation 
\begin{align}
	&\log_{2}\big(\varsigma_{n}^{(i)}\big) +\frac{\varsigma_{n}-\varsigma_{n}^{(i)}}{\ln (2) \varsigma_{n}^{(i)}}  \leq  \alpha^p_n, \forall n\in \mathcal N \label{43}
\end{align}
where $\varsigma_{n}^{(i)}$ denotes the value of $\varsigma_{n}$ at the $i$-th iteration.

Next, to trace the convexification of \eqref{36c}, we introduce a vector of slack variables $\boldsymbol{\gamma}_{n}=[\gamma_{n,1},\gamma_{n,2}]^T$ as follows:
	\begin{align}
	&  \log_{2}\bigl( \gamma_{n,1} \bigr) - \log_{2}\bigl(\gamma_{n,2}\bigr) \geq \sum\limits_{{\substack{n=1}}}^N  {r}^c_n,    \label{44}
\end{align}
	\begin{align}
		&\eta_n\operatorname{tr}\Big(\Big({\mathbf{F}}_{n}-\mu_n^2 \mathbf{I}\Big)\mathbf{B}_c \Big)+ \eta_n\operatorname{tr}\!\Big(
		\Big({\mathbf{F}}_{n}+\mu_n^2 \mathbf{I}\Big) \boldsymbol{\Psi}_1
		\Big)\nonumber \\
		&+ {\sigma}_n^2 \geq \gamma_{n,1}, \forall n\in \mathcal N,\label{45} 
	\end{align}
	\begin{align}
		& \eta_n\operatorname{tr}\!\Big(
		\Big({\mathbf{F}}_{n}+\mu_{n}^2 \mathbf{I}\Big) \boldsymbol{\Psi}_1
		\Big)
		+ {\sigma}_n^2  \leq \gamma_{n,2}, \forall n\in \mathcal N, \label{46} 
	\end{align}

To obtain a tractable reformulation of \eqref{44}, the second logarithmic term is substituted with its first-order upper-bound approximation given as
\begin{align}
	& \log_{2}\left(\gamma_{n,1}\right)
	-\left[
	\log_{2}\left(\gamma_{n,2}^{(i)}\right)
	+
	\frac{\gamma_{n,2}-\gamma_{n,2}^{(i)}}
	{\ln(2)\gamma_{n,2}^{(i)}}
	\right]\geq \sum\limits_{{\substack{n=1}}}^N  {r}^c_n, \label{47} 
\end{align}
where $\gamma_{n,2}^{(i)}$ denotes the value of ${\gamma_{n,2}}$ at the $i$-th iteration.

Finally, problem $\textbf{(P3)}$ can be equivalently reformulated as follows:
\begin{subequations}\label{P4}
	\begin{align}
		\text{\textbf{(P4)}}	& \max_{\substack{
				\mathbf B_c,\{\mathbf B_n,\boldsymbol{\gamma}_n,r_n^c,\tau_n^p,\alpha_n^p,\varsigma_n\}_{n\in\mathcal N}
		}} \  \sum\limits_{{\substack{n=1}}}^N \Big(r^c_n + \tau_n^p\Big), \label{48a}\\
		s.t.\ \ &  \eqref{21k},\eqref{34}, \eqref{36d},\eqref{36g}, \eqref{36h}, \eqref{38a}, \eqref{39}, \eqref{42}, \eqref{43}, \nonumber \\
		& \eqref{45},\eqref{46}, \eqref{47}, \label{48b}\\
		\ \ & \mathbf{B}_c \succeq 0, \ \mathbf{B}_n \succeq 0, \ \forall n\in \mathcal N, \label{48c}\\
		\ \ & \operatorname{rank}(\mathbf{B}_c)=1, \ \operatorname{rank}(\mathbf{B}_n)=1,  \forall n\in \mathcal N. \label{48d}
	\end{align}
\end{subequations}

To obtain a tractable formulation of \textbf{(P4)}, the rank-one constraints in \eqref{48d} are relaxed. Rank-one solutions directly yield the common- and private-stream precoders through eigenvalue decomposition. When higher-rank matrices are obtained, candidate precoders are constructed via Gaussian randomization \cite{ni2021resource}. Their associated common-rate allocations and precoding vectors are checked against the constraints of \textbf{(P3)}. Among the feasible candidates and the preceding design, the configuration attaining the largest sum-rate is adopted. This selection maintains feasibility and guarantees a nondecreasing sum-rate for the active-beamforming update.

\subsection{Updating RIS Configuration}
First, let us define a PSD matrix $\mathbf E=\mathbf e\mathbf e^H$, satisfying $\mathbf E\succeq\mathbf0$ and $\operatorname{rank}(\mathbf E)=1$, where $\mathbf e=[e^{j\varphi_1},e^{j\varphi_2},\ldots,e^{j\varphi_Q}]^T$. Since the RIS coefficients are optimized in this subproblem, the uncertainty radius varies with the RIS configuration. Let $\mathbf D_n\triangleq
\operatorname{diag}(\mathbf h_{r,n}^{H})\mathbf H_{b,r}$. Using $\mathbf f_n^H=\mathbf e^T\mathbf D_n$,
the uncertainty radius is expressed as $\mu_n^2(\mathbf E)=\mu\|\mathbf f_n\|_2^2 =\mu\operatorname{tr}\!\left( (\mathbf D_n\mathbf D_n^H)^T\mathbf E\right)$. Then, the rate expressions in \eqref{32} and \eqref{33} can be updated as
\begin{equation}
	\overline R_n^{c}
	=
	\log_2\!\left(
	1+
	\frac{
		\eta_n\operatorname{tr}(\mathbf E\mathbf G_1)
	}{
		\eta_n\operatorname{tr}(\mathbf E\mathbf G_2)
		+\sigma_n^2
	}
	\right), \label{49}
\end{equation}
and
\begin{equation}
	\overline R_n^{p}
	=
	\log_2\!\left(
	1+
	\frac{
		\eta_n\operatorname{tr}(\mathbf E\mathbf C_1)
	}{
		\eta_n\operatorname{tr}(\mathbf E\mathbf C_2)
		+\sigma_n^2
	}
	\right), 	\label{50}
\end{equation}
where
\begin{equation*}
	\begin{aligned}
		\mathbf G_1
		&=
		\left(
		\mathbf D_n
		[\mathbf B_c-\mu\operatorname{tr}(\mathbf B_c)\mathbf I_K]
		\mathbf D_n^H
		\right)^T,\\
		\mathbf G_2
		&=
		\left(
		\mathbf D_n
		[\boldsymbol\Psi_1
		+\mu\operatorname{tr}(\boldsymbol\Psi_1)\mathbf I_K]
		\mathbf D_n^H
		\right)^T,\\
		\mathbf C_1
		&=
		\left(
		\mathbf D_n
		[\mathbf B_n-\mu\operatorname{tr}(\mathbf B_n)\mathbf I_K]
		\mathbf D_n^H
		\right)^T,\\
		\mathbf C_2
		&=
		\left(
		\mathbf D_n
		[\boldsymbol\Psi_2
		+\mu\operatorname{tr}(\boldsymbol\Psi_2)\mathbf I_K]
		\mathbf D_n^H
		\right)^T.
	\end{aligned}
\end{equation*}

Accordingly, \eqref{34} can be reformulated as
\begin{equation}
	(1-\eta_n)\operatorname{tr}(\mathbf E\mathbf G_3)
	+\hat{\sigma}_n^2
	\geq A_n,
	\qquad \forall n\in\mathcal N, 	\label{51}
\end{equation}
where
$\mathbf G_3=
\left(
\mathbf D_n
\left[
\boldsymbol\Psi_{\rm EH}
-\mu\operatorname{tr}(\boldsymbol\Psi_{\rm EH})
\mathbf I_K
\right]
\mathbf D_n^H
\right)^T$.

Next, for given precoding vectors, PS ratios, and MAs positions, the original optimization problem reduces to the following RIS scattering matrix design subproblem:
\begin{subequations}\label{P5}
	\begin{align}
		\text{\textbf{(P5)}}	& \max_{\mathbf E,\{r_n^c\}_{n\in\mathcal N}} \    \sum\limits_{{\substack{n=1}}}^N \Big({r}^c_n + \overline{R}^p_{n}\Big), \label{52a}  \\
		s.t.\ \ & {R}_{\min} \leq {r}^c_n + \overline{R}^p_{n}, \forall n\in \mathcal N, \label{52b} \\
		\ \ & \sum\limits_{{\substack{n=1}}}^N  {r}^c_n  \leq \overline{R}^c_{n}, \forall n\in \mathcal N, \label{52c}\\
		\ \ & [\mathbf{E}]_{q,q} = 1, \forall q\in \mathcal Q, \label{52d}\\
		\ \ & {\mathbf{E}} \succeq \mathbf{0}, \label{52e}\\
		\ \ & \operatorname{rank}({\mathbf{E}})=1, \label{52f} \\
		\ \ & \operatorname{tr}(\mathbf E\mathbf G_1)\geq 0, \ \operatorname{tr}(\mathbf E\mathbf C_1)\geq 0, \qquad \forall n\in\mathcal N, \label{52g} \\
		\ \ & \eqref{21k},\eqref{51}. \label{52h}
	\end{align}
\end{subequations}

Problem \textbf{(P5)} remains non-convex due to its non-convex objective function and constraints. To facilitate the subsequent reformulation into a tractable form, the auxiliary variable vector $\boldsymbol{\chi}_{n}=[\chi_{n,1},\chi_{n,2}]^T$ is introduced as
\begin{subequations}
	\begin{align}
		 {R}_{\min} \leq {r}^c_n + \big(\chi_{n,1} - \chi_{n,2} \big), \forall n\in \mathcal N, \label{53a}  
	\end{align}
\begin{equation}
		\scalemath{0.95}{\log_2\!\left(
		\eta_n\operatorname{tr}
		\big(\mathbf E(\mathbf C_1+\mathbf C_2)\big)
		+\sigma_n^2
		\right)
		\geq \chi_{n,1},
		\qquad \forall n\in\mathcal N.}  \label{53b}
\end{equation}
\begin{equation}
	\log_2\!\left(
	\eta_n\operatorname{tr}(\mathbf E\mathbf C_2)
	+\sigma_n^2
	\right)
	\leq \chi_{n,2},
	\qquad \forall n\in\mathcal N.  \label{53c}
\end{equation}
\end{subequations}

Further, the non-convex constraint in \eqref{52c} can be equivalently reformulated by introducing the auxiliary variable vector $\boldsymbol{\upsilon}_{n}=[\upsilon_{n,1},\upsilon_{n,2}]^T$ as
\begin{subequations}
	\begin{align}
		\sum\limits_{{\substack{n=1}}}^N  {r}^c_n \leq  \big(\upsilon_{n,1} - \upsilon_{n,2} \big), \forall n\in \mathcal N, \label{54a}  
	\end{align}
	\begin{equation}
			\scalemath{0.95}{\log_2\!\left(
		\eta_n\operatorname{tr}
		\big(\mathbf E(\mathbf G_1+\mathbf G_2)\big)
		+\sigma_n^2
		\right)
		\geq \upsilon_{n,1},
		\qquad \forall n\in\mathcal N,}\label{54b}
	\end{equation}
	\begin{equation}
		\log_2\!\left(
		\eta_n\operatorname{tr}(\mathbf E\mathbf G_2)
		+\sigma_n^2
		\right)
		\leq \upsilon_{n,2},
		\qquad \forall n\in\mathcal N, \label{54c}
	\end{equation}
\end{subequations}

Subsequently, to handle the remaining non-convexity in \eqref{53c} and \eqref{54c}, we introduce a vector of slack variables $\boldsymbol{\varrho}_{n}=[\varrho_{n,1},\varrho_{n,2}]^T$ defined as:
\begin{subequations}
		\begin{align}
		&\log_{2}\big( \varrho_{n,1}\big) \leq \chi_{n,2}, \forall n\in \mathcal N , \label{55a}  
	\end{align}
	\begin{align}
		&\log_{2}\big( \varrho_{n,2}\big) \leq \upsilon_{n,2}, \forall n\in \mathcal N , \label{55b}  
	\end{align}
	\begin{equation}
		\eta_n\operatorname{tr}(\mathbf E\mathbf C_2)
		+\sigma_n^2
		\leq \varrho_{n,1},
		\qquad \forall n\in\mathcal N, \label{55c}
	\end{equation}
	\begin{equation}
		\eta_n\operatorname{tr}(\mathbf E\mathbf G_2)
		+\sigma_n^2
		\leq \varrho_{n,2},
		\qquad \forall n\in\mathcal N, \label{55d}
	\end{equation}
\end{subequations}
 
 Since \eqref{55a} and \eqref{55b} remain non-convex, they are replaced by their corresponding convex surrogate functions, given as
\begin{align}
	&\log_{2}\big(\varrho_{n,1}^{(i)}\big) +\frac{\varrho_{n,1}-\varrho_{n,1}^{(i)}}{\ln (2) \varrho_{n,1}^{(i)}}  \leq \chi_{n,2}, \forall n\in \mathcal N \label{56}
\end{align}
\begin{align}
	&\log_{2}\big(\varrho_{n,2}^{(i)}\big) +\frac{\varrho_{n,2}-\varrho_{n,2}^{(i)}}{\ln (2) \varrho_{n,2}^{(i)}}  \leq \upsilon_{n,2}, \forall n\in \mathcal N  \label{57}
\end{align}
Based on the above transformations, problem \textbf{(P5)} is equivalently reformulated as
\begin{subequations}\label{P6}
	\begin{align}
		\text{\textbf{(P6)}}	& \max_{\substack{\mathbf E,\{r_n^c,\boldsymbol{\chi}_n, \boldsymbol{\upsilon}_n,\boldsymbol{\varrho}_n\}_{n\in\mathcal N}}} \  \sum\limits_{{\substack{n=1}}}^N {r}^c_n + \big(\chi_{n,1} - \chi_{n,2} \big), \label{58a}  \\
		s.t. \ \ & \eqref{21k},\eqref{51}, \eqref{52g}, \eqref{53a}, \eqref{53b}, \eqref{54a}, \eqref{54b}, \eqref{55c}, \nonumber \\
		& \eqref{55d},\eqref{56}, \eqref{57}, \label{58b} \\  
	    \ \ & [\mathbf{E}]_{q,q} = 1, \forall q\in \mathcal Q, \label{58c}\\
	    \ \ & {\mathbf{E}} \succeq \mathbf{0}, \label{58d}\\
	    \ \ & \operatorname{rank}({\mathbf{E}})=1 \label{58e} .    
	\end{align}
\end{subequations}
where the rank-one constraint in \eqref{58e} renders \textbf{(P6)} non-convex. By relaxing this constraint, \textbf{(P6)} can be reformulated as a tractable convex optimization problem. If the resulting solution is rank one, the RIS coefficient vector is recovered by eigenvalue decomposition;
otherwise, Gaussian randomization is employed to generate rank-one candidates \cite{ni2021resource}. Additionally, each candidate, together with its associated common-rate allocation, is evaluated using the objective and constraints of \textbf{(P5)}. The previous feasible design is retained as a candidate, and the feasible design achieving the highest sum-rate is selected. Thus, the passive-beamforming update preserves feasibility and ensures a nondecreasing sum-rate.

\subsection{MAs Position Optimization}
For fixed transmit precoders, RIS coefficients, and PS ratios, the MAs positions and common-rate allocations are optimized as
\begin{subequations}\label{P7}
	\begin{align}
		\text{\textbf{(P7)}}\quad
		& \max_{\widehat{\boldsymbol{\zeta}},
			\{r_n^c\}_{n\in\mathcal N}}
		\sum_{n=1}^{N}(r_n^c+\overline R_n^p)
		\label{59a}\\
		\mathrm{s.t.}\quad
		&\eqref{21i},\ \eqref{34},\
		\eqref{52b},\ \eqref{52c},
		\label{59b}\\
		&\|\widehat{\boldsymbol{\zeta}}_k-
		\widehat{\boldsymbol{\zeta}}_j\|_2\geq D_{\min},
		\quad \forall k,j\in\mathcal K,\quad k\neq j,
		\label{59c}\\
		&r_n^c\geq0,\quad \forall n\in\mathcal N.
		\label{eq:ma_common_nonnegative}
	\end{align}
\end{subequations}
The nonlinear position-dependent channels and coupled spacing constraints make \textbf{(P7)} non-convex. We update one antenna position at a time, keeping the remaining positions fixed. For antenna $k$, let $\mathbf z=\widehat{\boldsymbol{\zeta}}_k$, $\mathbf z^{(i)}=\widehat{\boldsymbol{\zeta}}_k^{(i)}$, and $\Delta\mathbf z=\mathbf z-\mathbf z^{(i)}$. The update region $\mathcal C_k^{(i)}$ is compact, convex, contains $\mathbf z^{(i)}$, and lies within the permitted movement region. It is chosen so that the rate denominators and logarithm arguments remain strictly positive and the functions below are twice continuously differentiable on its neighborhood.

\medskip
\noindent\textbf{Lemma 1:}
Let $f$ be twice continuously differentiable on a neighborhood of a convex set $\mathcal C$.
If $L_f\geq0$ satisfies
\begin{equation}
	-L_f\mathbf I\preceq\nabla^2f(\mathbf z)
	\preceq L_f\mathbf I,\quad \forall\mathbf z\in\mathcal C,
	\label{eq:ma_curvature_condition}
\end{equation}
define
$T_f^{(i)}(\mathbf z)
=f(\mathbf z^{(i)})+
\nabla f(\mathbf z^{(i)})^T\Delta\mathbf z$.
Then
\begin{align}
	\underline f^{(i)}(\mathbf z)
	&=T_f^{(i)}(\mathbf z)
	-\tfrac{L_f}{2}\|\Delta\mathbf z\|_2^2,
	\label{eq:ma_general_lower}\\
	\overline f^{(i)}(\mathbf z)
	&=T_f^{(i)}(\mathbf z)
	+\tfrac{L_f}{2}\|\Delta\mathbf z\|_2^2
	\label{eq:ma_general_upper}
\end{align}
satisfy $\underline f^{(i)}(\mathbf z)\leq f(\mathbf z) \leq\overline f^{(i)}(\mathbf z)$ on $\mathcal C$.
Both bounds match the value and gradient of $f$ at $\mathbf z^{(i)}$. A sufficient choice is $L_f\geq\sup_{\mathbf z\in\mathcal C} \|\nabla^2f(\mathbf z)\|_2$; the Hessian norm at the current point alone is insufficient. Accordingly, with the remaining variables fixed, denote the common and private rates by $\overline R_n^s(\mathbf z)$, $s\in\{c,p\}$. Additionally, the uncertainty radius is evaluated as $\mu_n^2(\mathbf z)=\mu\|\mathbf f_n(\mathbf z)\|_2^2$, and its position dependence is included in the gradients and curvature bounds of the rate and EH functions. 

Let $\overline R_n^{s(i)} =\overline R_n^s(\mathbf z^{(i)})$ and $\mathbf g_{n,k}^{s(i)} =\nabla_{\mathbf z}\overline R_n^s(\mathbf z^{(i)})$. Applying the lemma 1 gives
\begin{equation}
	\underline R_{n,k}^{s(i)}(\mathbf z)
	=\overline R_n^{s(i)}
	+(\mathbf g_{n,k}^{s(i)})^T\Delta\mathbf z
	-\tfrac{L_{n,k}^{s(i)}}{2}\|\Delta\mathbf z\|_2^2,
	\label{eq:ma_rate_lower}
\end{equation}
where $L_{n,k}^{s(i)}$ satisfies \eqref{eq:ma_curvature_condition} for $\overline R_n^s$ on $\mathcal C_k^{(i)}$. These concave surrogates lower-bound the rates and are value- and gradient-matched at the current position.

Next, we introduce the user-rate slack vector $\widehat{\boldsymbol{\lambda}}^p =[\widehat\lambda_1^p,\ldots,\widehat\lambda_N^p]^T$ such that
\begin{equation}
	\widehat\lambda_n^p
	\leq\overline R_n^p(\mathbf z),
	\quad \forall n\in\mathcal N.
	\label{60}
\end{equation}
Since the quadratic surrogate satisfies $\underline R_{n,k}^{p(i)}(\mathbf z) \leq\overline R_n^p(\mathbf z)$, replacing the rate in \eqref{60} by this lower bound yields the sufficient convex constraint
\begin{equation}
	\widehat\lambda_n^p
	\leq\underline R_{n,k}^{p(i)}(\mathbf z),
	\quad \forall n\in\mathcal N.
	\label{61}
\end{equation}
Similarly, using $\underline R_{n,k}^{c(i)}(\mathbf z) \leq\overline R_n^c(\mathbf z)$, the common-stream decoding constraint \eqref{52c} is conservatively approximated as
\begin{equation}
	\sum_{m=1}^{N}r_m^c
	\leq\underline R_{n,k}^{c(i)}(\mathbf z),
	\quad \forall n\in\mathcal N.
	\label{62}
\end{equation}

To preserve the EH constraint, let $\mathcal E_{n,k}(\mathbf z)$ denote the left-hand side of \eqref{34} with the remaining variables fixed. Define $E_{n,k}^{(i)}=\mathcal E_{n,k}(\mathbf z^{(i)})$ and
$\mathbf g_{n,k}^{E(i)} =\nabla_{\mathbf z}\mathcal E_{n,k}(\mathbf z^{(i)})$. The lemma 1 yields
\begin{equation}
	\underline{\mathcal E}_{n,k}^{(i)}(\mathbf z)
	=E_{n,k}^{(i)}
	+(\mathbf g_{n,k}^{E(i)})^T\Delta\mathbf z
	-\tfrac{L_{n,k}^{E(i)}}{2}\|\Delta\mathbf z\|_2^2,
	\label{eq:ma_eh_lower}
\end{equation}
where $L_{n,k}^{E(i)}$ satisfies \eqref{eq:ma_curvature_condition} for $\mathcal E_{n,k}$ on $\mathcal C_k^{(i)}$. Thus, a conservative EH constraint is
\begin{equation}
	\underline{\mathcal E}_{n,k}^{(i)}(\mathbf z)
	\geq A_n,\quad \forall n\in\mathcal N.
	\label{eq:ma_eh_constraint}
\end{equation}

Accordingly, let $\underline S_{n,k}^{s(i)}(\mathbf z)$ denote the concave lower bound obtained from Lemma~1 for $S_n^s(\mathbf z) =\operatorname{tr}[(\mathbf F_n(\mathbf z)-\mu_n^2(\mathbf z)\mathbf I)\mathbf B_{n,s}]$, where $\mathbf B_{n,c}=\mathbf B_c$ and $\mathbf B_{n,p}=\mathbf B_n$.

Further, for spacing constraint, define the fixed unit vector $\mathbf d_{kj}^{(i)} =(\mathbf z^{(i)}-\widehat{\boldsymbol{\zeta}}_j)/ \|\mathbf z^{(i)}-\widehat{\boldsymbol{\zeta}}_j\|_2$. Convexity of the Euclidean norm gives
\begin{equation}
	\|\mathbf z-\widehat{\boldsymbol{\zeta}}_j\|_2
	\geq(\mathbf d_{kj}^{(i)})^T
	(\mathbf z-\widehat{\boldsymbol{\zeta}}_j).
	\label{63}
\end{equation}

The resulting single-antenna position-update subproblem is
\begin{subequations}\label{P8}
	\begin{align}
		\text{\textbf{(P8)}}\quad
		& \max_{\mathbf z,\widehat{\boldsymbol{\lambda}}^p,
			\{r_n^c\}_{n\in\mathcal N}}
		\sum_{n=1}^{N}(r_n^c+\widehat\lambda_n^p)
		\label{70a}\\
		\mathrm{s.t.}\quad
		&r_n^c+\widehat\lambda_n^p\geq R_{\min},
		\quad \forall n\in\mathcal N,
		\label{70b}\\
		&(\mathbf d_{kj}^{(i)})^T
		(\mathbf z-\widehat{\boldsymbol{\zeta}}_j)
		\geq D_{\min},
		\quad \forall j\in\mathcal K\setminus\{k\},
		\label{70c}\\
		&\eqref{61},\ \eqref{62},\
		\eqref{eq:ma_eh_constraint},
		\label{70d}\\
		&\mathbf z\in\mathcal C_k^{(i)},\quad
		r_n^c\geq0,\quad \forall n\in\mathcal N,
		\label{70e} \\
		& \underline S_{n,k}^{s(i)}(\mathbf z)\geq0,
		\qquad s\in\{c,p\},\ \forall n\in\mathcal N. \label{70f}
	\end{align}
\end{subequations}
Problem \textbf{(P8)} is convex and can be solved using CVX with MOSEK. The bounds are tight at the current feasible position, ensuring that the current design, with $\widehat\lambda_n^p=\overline R_n^{p(i)}$, remains feasible for \textbf{(P8)}. Since the surrogate rates lower-bound the original rates, solving \textbf{(P8)} to optimality yields a feasible position and common-rate update with a
nondecreasing sum-rate. Antenna positions are updated sequentially until the stopping criterion is met.

\subsection{Optimization of PS ratios}
For fixed precoding matrices, RIS coefficients, and MAs positions, the PS and common-rate allocation
subproblem is
\begin{subequations}\label{P9}
	\begin{align}
		\text{\textbf{(P9)}}\quad
		&\max_{\{r_n^c,\eta_n\}_{n\in\mathcal N}}
		\sum_{n=1}^{N}\left(r_n^c+\overline R_n^p\right)
		\label{71a},\\
		\text{s.t.}\quad
		&\eqref{34},\ \eqref{36b},\ \eqref{36c},
		\label{71b}\\
		&r_n^c\geq0,\quad 0\leq\eta_n\leq1,
		\quad \forall n\in\mathcal N.
		\label{71c}
	\end{align}
\end{subequations}
Define
\begin{equation}
	C_n\triangleq
	\operatorname{tr}\!\left[
	(\mathbf F_n-\mu_n^2\mathbf I)
	\boldsymbol\Psi_{\rm EH}
	\right]
	+\bar{\sigma}_n^2.
	\label{eq:ps_received_coefficient}
\end{equation}
Using $\hat{\sigma}_n^2=(1-\eta_n)\bar{\sigma}_n^2$, constraint~\eqref{34} becomes
\begin{equation}
	(1-\eta_n)C_n\geq A_n,
	\quad \forall n\in\mathcal N,
	\label{eq:ps_eh_compact}
\end{equation}
where $A_n$ is given in~\eqref{35}.

\noindent\textbf{Proposition 1
(Closed-form PS and common-rate allocation):}
Suppose $C_n>0$, $0\leq A_n\leq C_n$, $\widetilde{\sigma}_{\mathrm d}^2>0$, and the desired-signal coefficients in~\eqref{32} and~\eqref{33} are nonnegative.
Define
\begin{align}
	\eta_n^{\max}
	&\triangleq1-\frac{A_n}{C_n},
	\label{eq:ps_upper_bound}\\
	d_n
	&\triangleq
	\max\!\left\{
	0,R_{\min}-\overline R_n^p(\eta_n^{\max})
	\right\},
	\label{eq:ps_rate_deficit}\\
	R_c^{\mathrm{avail}}
	&\triangleq
	\min_{n\in\mathcal N}
	\overline R_n^c(\eta_n^{\max}),
	\label{eq:ps_available_common_rate}
\end{align}
where the rates are evaluated at $\eta_n=\eta_n^{\max}$. Then \textbf{(P9)} is feasible if and only if
\begin{equation}
	\sum_{n=1}^{N}d_n\leq R_c^{\mathrm{avail}}.
	\label{eq:ps_rate_feasibility}
\end{equation}
When feasible, an optimal solution is
\begin{align}
	\eta_n^\star&=\eta_n^{\max},
	\label{eq:ps_closed_form}\\
	r_n^{c\star}
	&=d_n+
	\frac{R_c^{\mathrm{avail}}-\sum_{m=1}^{N}d_m}{N},
	\quad \forall n\in\mathcal N.
	\label{eq:ps_common_allocation}
\end{align}

\begin{proof}
For fixed remaining variables, \eqref{32} and~\eqref{33} have the form
	\begin{equation}
		\overline R_n^s(\eta_n)
		=
		\log_2\!\left(
		1+\frac{\eta_n a_n^s}
		{\eta_n b_n^s+\widetilde{\sigma}_{\mathrm d}^2}
		\right),
		\quad s\in\{c,p\},
		\label{eq:ps_rate_compact}
	\end{equation}
	where $a_n^c=\operatorname{tr}[(\mathbf F_n-\mu_n^2\mathbf I)\mathbf B_c]$ and $a_n^p=\operatorname{tr}[ (\mathbf F_n-\mu_n^2\mathbf I)\mathbf B_n]$. The coefficients $b_n^c$ and $b_n^p$ are the corresponding interference-and-distortion trace terms in~\eqref{32} and~\eqref{33}, respectively, plus $(1+\varrho_r)\bar{\sigma}_n^2$. These coefficients are independent of $\eta_n$
	and satisfy $a_n^s,b_n^s\geq0$. Since
	\begin{equation}
		\frac{\partial}{\partial\eta_n}
		\left(
		\frac{\eta_n a_n^s}
		{\eta_n b_n^s+\widetilde{\sigma}_{\mathrm d}^2}
		\right)
		=
		\frac{a_n^s\widetilde{\sigma}_{\mathrm d}^2}
		{(\eta_n b_n^s+\widetilde{\sigma}_{\mathrm d}^2)^2}
		\geq0,
		\label{eq:ps_sinr_monotonicity}
	\end{equation}
	both rates are nondecreasing in $\eta_n$.
	
	Constraint~\eqref{eq:ps_eh_compact} imposes $\eta_n\leq\eta_n^{\max}$. Increasing any feasible PS ratios to these upper bounds preserves the EH, QoS, and common-stream decoding constraints and
	cannot decrease the objective. It therefore suffices to consider $\eta_n=\eta_n^{\max}$.
	
	At these ratios, the rate constraints reduce to $r_n^c\geq d_n$ and $\sum_n r_n^c\leq R_c^{\mathrm{avail}}$, which are feasible exactly when \eqref{eq:ps_rate_feasibility} holds. The objective is maximized by allocating all available common rate. Equation~\eqref{eq:ps_common_allocation} satisfies these constraints and $\sum_n r_n^{c\star}=R_c^{\mathrm{avail}}$, completing the proof.
\end{proof}

\subsection{Analytical Results and Design Insights}

\subsubsection{High-Power Behavior and Hardware-Impairment Limit}

Consider the SINRs in~\eqref{14} and~\eqref{15} for fixed RIS coefficients, MAs positions, ctual effective channel, PS ratios $\eta_n>0$, and noise powers. Let $\mathbf b_c=\sqrt{P}\,\widetilde{\mathbf b}_c$ and $\mathbf b_n=\sqrt{P}\,\widetilde{\mathbf b}_n$, where $P>0$ denotes the total intended transmit power. The fixed normalized precoders satisfy
$\|\widetilde{\mathbf b}_c\|_2^2+
\sum_{j=1}^{N}\|\widetilde{\mathbf b}_j\|_2^2=1$.
Define
$\widetilde{\mathbf W}
=\widetilde{\mathbf b}_c\widetilde{\mathbf b}_c^H+
\sum_{j=1}^{N}
\widetilde{\mathbf b}_j\widetilde{\mathbf b}_j^H$.

To describe both decoding stages, let $s\in\{c,p\}$ and define
\begin{equation}
	\widetilde{\mathbf v}_{n,s}
	=
	\begin{cases}
		\widetilde{\mathbf b}_c, & s=c,\\
		\widetilde{\mathbf b}_n, & s=p,
	\end{cases}
	\qquad
	\mathcal J_{n,s}
	=
	\begin{cases}
		\mathcal N, & s=c,\\
		\mathcal N\setminus\{n\}, & s=p.
	\end{cases}
	\label{eq:hp_decoding_definitions}
\end{equation}
Thus, all private streams interfere with common-stream decoding, whereas private-stream decoding experiences interference from the other users' private streams and residual common-stream interference due to imperfect SIC.

For the actual effective-channel outer-product matrix $\overline{\mathbf F}_n=\overline{\mathbf f}_n\overline{\mathbf f}_n^H=\mathbf F_n+\Delta\mathbf F_n\succeq\mathbf0$, 
define
\begin{align}
	S_n^s
	&\triangleq
	\operatorname{tr}\!\left(
	\overline{\mathbf F}_n
	\widetilde{\mathbf v}_{n,s}
	\widetilde{\mathbf v}_{n,s}^{H}
	\right),
	\label{eq:hp_signal_coefficient}\\
	D_n^s
	&\triangleq
	\sum_{j\in\mathcal J_{n,s}}
	\operatorname{tr}\!\left(
	\overline{\mathbf F}_n
	\widetilde{\mathbf b}_j\widetilde{\mathbf b}_j^H
	\right)
	+\varrho_r
	\operatorname{tr}\!\left(
	\overline{\mathbf F}_n\widetilde{\mathbf W}
	\right)
	\nonumber\\
	&\quad+
	(1+\varrho_r)\varrho_t
	\operatorname{tr}\!\left(
	\overline{\mathbf F}_n
	\widetilde{\operatorname{diag}}
	(\widetilde{\mathbf W})
	\right)
	\nonumber\\
	&\quad+
	\delta_n\mathbf{1}_{\{s=p\}}
	\operatorname{tr}\!\left(
	\overline{\mathbf F}_n
	\widetilde{\mathbf b}_c\widetilde{\mathbf b}_c^H
	\right).
	\label{eq:hp_distortion_coefficient}
\end{align}
Here, $\mathbf{1}_{\{s=p\}}$ equals one for private-stream
decoding and zero for common-stream decoding.

\noindent\textbf{Proposition 2 (High-power SINR limits):} For the fixed configuration above, if $S_n^s>0$
and $D_n^s>0$, then
\begin{equation}
	\lim_{P\to\infty}\overline{\Gamma}_n^s(P)
	=
	\frac{S_n^s}{D_n^s},
	\qquad s\in\{c,p\}.
	\label{eq:hp_sinr_limit}
\end{equation}
The corresponding decoding rate approaches
$\log_2(1+S_n^s/D_n^s)$.

\begin{proof}
	Substituting the scaled precoders into \eqref{14} and~\eqref{15} gives
	$\overline{\Gamma}_n^s(P)
	=\eta_n P S_n^s/(\eta_n P D_n^s+\sigma_n^2)$.
	Using
	$\sigma_n^2
	=\eta_n(1+\varrho_r)\bar{\sigma}_n^2+
	\widetilde{\sigma}_{\mathrm d}^2$
	yields
	\begin{equation}
		\overline{\Gamma}_n^s(P)
		=
		\frac{S_n^s}{
			D_n^s+
			\frac{(1+\varrho_r)\bar{\sigma}_n^2}{P}
			+
			\frac{\widetilde{\sigma}_{\mathrm d}^2}{\eta_n P}
		}.
		\label{eq:hp_normalized_sinr}
	\end{equation}
	Since $\eta_n>0$ is fixed, both noise terms vanish as $P\to\infty$, establishing \eqref{eq:hp_sinr_limit}. The rate limit follows from the continuity of $\log_2(1+x)$.
\end{proof}

\noindent\textbf{Corollary 1 (Receiver-distortion bound).}
If $\varrho_r>0$, $\varrho_t\geq0$, and $S_n^s>0$, then, for every $P>0$,
\begin{equation}
	\overline{\Gamma}_n^s(P)
	\leq
	\frac{S_n^s}{D_n^s}
	\leq
	\frac{1}{\varrho_r},
	\qquad s\in\{c,p\}.
	\label{eq:receiver_distortion_bound}
\end{equation}
Each corresponding decoding rate is therefore upper bounded by $\log_2(1+1/\varrho_r)$.

\subsubsection{Single-Path Limit of Antenna-Position Optimization}

\noindent\textbf{Proposition 3
(Single-transmit-path equivalence):}
Under the channel model in~\eqref{2}--\eqref{7}, suppose that $K_t=1$, the path gains and angles remain fixed over the movement region, and the complex precoding coefficients are freely adjustable. The channel uncertainty follows the bounded matrix-error model $\|\Delta\mathbf F_n\|_2\leq\mu\|\mathbf F_n\|_2$. Then, under the adopted transceiver-distortion model, any two feasible MAs layouts yield the same optimal sum-rate after precoder adjustment.

\begin{proof}
	For $K_t=1$, equations~\eqref{2}--\eqref{7} give
	\begin{equation}
		\mathbf f_n^H(\widehat{\boldsymbol{\zeta}})
		=
		c_n
		\begin{bmatrix}
			e^{j\psi_1}&\cdots&e^{j\psi_K}
		\end{bmatrix},
		\label{eq:sp_effective_channel}
	\end{equation}
	where
	$\psi_k=\frac{2\pi}{\lambda}
	\Xi(\widehat{\boldsymbol{\zeta}}_k,\theta_1^e,\theta_1^a)$
	and
	$c_n=\mathbf h_{r,n}^H\boldsymbol{\Theta}
	\mathbf V^r(\widehat{\boldsymbol{\tau}})^H
	\boldsymbol{\Omega}$.
	For fixed RIS coefficients, $c_n$ is independent
	of the MAs positions.
	
	For two feasible layouts with phases $\{\psi_k\}$
	and $\{\psi_k'\}$, define
	$\mathbf U=\operatorname{diag}
	(e^{j(\psi_1-\psi_1')},\ldots,
	e^{j(\psi_K-\psi_K')})$.
	Then $\mathbf U^H\mathbf U=\mathbf I$ and
	$\mathbf f_n^H(\widehat{\boldsymbol{\zeta}}')
	=\mathbf f_n^H(\widehat{\boldsymbol{\zeta}})
	\mathbf U^H$.
	Choosing $\mathbf b_s'=\mathbf U\mathbf b_s$
	for $s\in\{c,1,\ldots,N\}$ yields
	\begin{equation}
		\mathbf f_n^H(\widehat{\boldsymbol{\zeta}}')
		\mathbf b_s'
		=
		\mathbf f_n^H(\widehat{\boldsymbol{\zeta}})
		\mathbf b_s,
		\qquad
		\|\mathbf b_s'\|_2^2=\|\mathbf b_s\|_2^2.
		\label{eq:sp_received_signal_invariance}
	\end{equation}
	Since the channel transformation preserves $\|\mathbf f_n\|_2^2$, it also preserves the
	uncertainty radius $\mu_n^2=\mu\|\mathbf f_n\|_2^2$. Thus, all received stream powers and the transmit power are preserved.
	
	For $\mathbf W=\mathbf b_c\mathbf b_c^H+ \sum_j\mathbf b_j\mathbf b_j^H$, the transformed covariance is $\mathbf W'=\mathbf U\mathbf W\mathbf U^H$, with $\widetilde{\operatorname{diag}}(\mathbf W') =\widetilde{\operatorname{diag}}(\mathbf W)$.
	Together with the channel transformation, this preserves the received transmitter- and receiver-distortion powers. Keeping the PS ratios and common-rate allocations unchanged therefore
	preserves the SINRs and harvested powers.
	
   The error transformation $\Delta\mathbf F_n'=\mathbf U\Delta\mathbf F_n\mathbf U^H$ preserves its spectral norm. Since the uncertainty radius is also preserved, the matrix uncertainty set and the robust constraints remain unchanged under this transformation. The inverse transformation provides the reverse mapping, establishing equality of the optimal objective values.
\end{proof}

\begin{algorithm}[t]
	\caption{Robust Resource Allocation Framework}
	\begin{algorithmic}[1]
		\State Initialize $\mathbf B_c$, $\{\mathbf B_n\}_{n\in\mathcal N}$, $\boldsymbol{\Theta}$, $\widehat{\boldsymbol{\zeta}}$, $\mathcal I_{\max}$, and auxiliary variables.
		\Repeat
		
		\State \textbf{Update precoding matrices:}
		\State Solve \textbf{(P4)} (see Section~III-A) to obtain $\mathbf B_c$ and $\{\mathbf B_n\}_{n\in\mathcal N}$.
		\If{$\operatorname{rank}(\mathbf{B}_c)=1$ and $\operatorname{rank}(\mathbf{B}_n)=1$}
		\State Recover $\mathbf b_c$ and $\{\mathbf b_n\}_{n\in\mathcal N}$ using eigenvalue decomposition.
		\Else
		\State Recover feasible rank-one solutions.
		\EndIf
		
		\State \textbf{Update RIS scattering matrix:}
		\State Solve \textbf{(P6)} (see Section~III-B) to obtain $\mathbf{E}$.
		\If{$\operatorname{rank}(\mathbf{E})=1$}
		\State Recover $\mathbf{e}$ through eigenvalue decomposition.
		\Else
		\State Recover a feasible rank-one solution.
		\EndIf
		
		\State \textbf{Update MAs positions:}
		\State Solve \textbf{(P8)} (see Section~III-C) to update $\widehat{\boldsymbol{\zeta}}$.
		
		\State \textbf{Update PS ratios:}
		\State Update $\{\eta_n,r_n^c\}_{n\in\mathcal N}$ using
		\eqref{eq:ps_closed_form} and \eqref{eq:ps_common_allocation}
		(Proposition~1, Section~III-D).
		
		\Until{convergence or $I_{\max}$ is reached.}
		
		\State \Return $\mathbf b_c^\star$, $\{\mathbf b_n^\star\}_{n\in\mathcal N}$, $\boldsymbol{\Theta}^\star$, $\widehat{\boldsymbol{\zeta}}^\star$, $\{\eta_n^\star\}_{n\in\mathcal N}$.
	\end{algorithmic}
\end{algorithm}

\subsection{Complexity and Convergence Analysis}    
\noindent \textit{1) Computational Complexity:} The primary computational complexity of Algorithm~1 arises from solving the transmit precoding, RIS scattering matrix, PS ratio, and MAs position optimization subproblems. Specifically, subproblems \textbf{(P4)} and \textbf{(P6)} are solved using the CVX framework based on interior-point methods, resulting in computational complexities of $\mathcal{O}\!\left(\tilde{\kappa}_1 K^{3.5}\right)$ and $\mathcal{O}\!\left(\tilde{\kappa}_2 Q^{3.5}\right)$ \cite{asif2026robust, wright1997primal}, respectively, where $\tilde{\kappa}_1$ and $\tilde{\kappa}_2$ represent the respective numbers of iterations required for \textbf{(P4)} and \textbf{(P6)} to converge. The PS-ratio subproblem (P9) is solved in closed form using Proposition~1, with a computational complexity of $\mathcal O(N)$. In addition, the MAs position optimization subproblem \textbf{(P8)} is solved through an iterative position-update process with computational complexity $\mathcal{O}\!\left(\widehat{\upsilon}\tilde{\kappa}_3 K\right)$, where $\widehat{\upsilon}$ denotes the outer iteration count required for the convergence of \textbf{(P8)}, and $\tilde{\kappa}_3$ denotes the number of iterations required for updating each MAs position. Since these four subproblems are successively solved within each outer iteration, the total computational complexity of Algorithm~1 is given as $\mathcal O\!\left(I_{\max}\left[\widetilde\kappa_1K^{3.5}+\widetilde\kappa_2Q^{3.5}
+N+\widehat\upsilon\widetilde\kappa_3 K\right]\right)$, where $I_{\max}$ represents the maximum number of outer iterations required for convergence.

\noindent\textit{2) Convergence Analysis:}
Let $\mathcal X$ collect the transmit precoders, RIS coefficients, MAs positions, PS ratios, and common-rate allocations, and let $\Psi(\mathcal X)$ denote the sum-rate objective. Starting from a
feasible initialization, Algorithm~1 successively updates four variable blocks.

For active and passive beamforming, the concavity of $\log_2(x)$ gives, for $x,x^{(i)}>0$,
\begin{equation}
	\log_2(x)\leq
	\log_2(x^{(i)})
	+\frac{x-x^{(i)}}{x^{(i)}\ln 2}.
	\label{eq:conv_log_bound}
\end{equation}
Applying this upper bound to the subtracted logarithmic terms yields rate lower bounds that match the original rates and their gradients at the current iterate. For MAs-position optimization, Lemma~1 provides concave quadratic lower bounds for the rates and EH expressions under the curvature condition in
\eqref{eq:ma_curvature_condition}. Together with the affine lower bound on antenna spacing, these approximations preserve the corresponding constraints and are tight at the current feasible position, although they may differ from the original functions at other feasible points.

Let $F$ denote a block objective and $\widetilde F$ its lower-bound surrogate, constructed at the current iterate. When the current design remains feasible for the surrogate subproblem, an optimal surrogate update satisfies
\begin{equation}
	F_{\mathrm{new}}
	\geq\widetilde F_{\mathrm{new}}
	\geq\widetilde F_{\mathrm{old}}
	=F_{\mathrm{old}}.
	\label{eq:conv_surrogate_ascent}
\end{equation}
This inequality applies before any rank-one recovery. In the active and passive beamforming subproblems, the rank relaxation is tight when an optimal solution satisfies the rank-one constraints, allowing lossless recovery. Otherwise, Gaussian randomization may introduce a loss relative to the objective value obtained before recovery. Selecting the best feasible design from the recovered candidates and the previous design ensures a nondecreasing sum-rate after recovery. Likewise, the closed-form PS and common-rate update preserves this property under the assumptions of Proposition~1.

Let $\mathcal X^{(i,\ell)}$ denote all design variables after the $\ell$-th block update within outer iteration $i$, with $\mathcal X^{(i,0)}=\mathcal X^{(i)}$ and $\mathcal X^{(i,4)}=\mathcal X^{(i+1)}$.
Combining the successive updates yields
\begin{equation}
	\Psi(\mathcal X^{(i+1)})
	\geq\Psi(\mathcal X^{(i,3)})
	\geq\Psi(\mathcal X^{(i,2)})
	\geq\Psi(\mathcal X^{(i,1)})
	\geq\Psi(\mathcal X^{(i)}).
	\label{70}
\end{equation}
Moreover, the finite transmit-power budget, bounded channel responses over the feasible movement region, and positive noise powers impose a finite upper limit on the achievable sum-rate. Consequently, the nondecreasing objective-value sequence generated by Algorithm~1 converges to a finite value.

\begin{figure}[!h]
	\centering
	\includegraphics [width=0.42\textwidth]{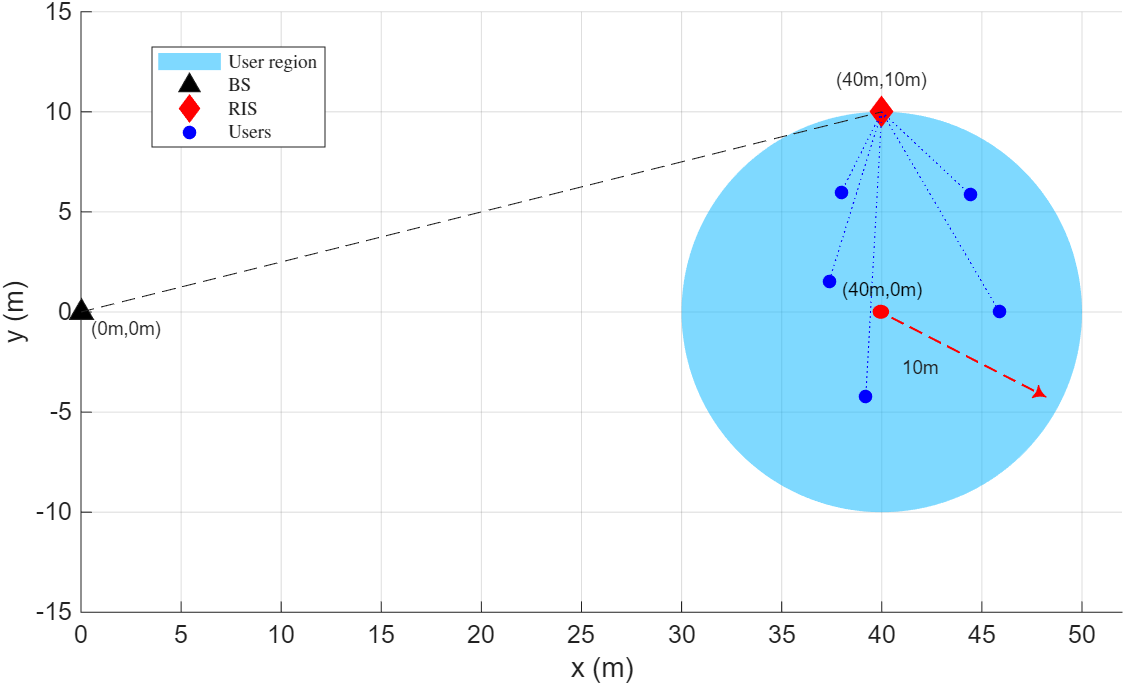}
	\caption{Simulation setup.}
	\label{f2}
\end{figure} 
\section{Numerical Results and Performance Analysis}
The proposed framework is evaluated through comprehensive Monte Carlo simulations using the network topology shown in Fig.~\ref{f2}. In this setup, the BS and RIS are deployed at $(0~\mathrm{m},0~\mathrm{m})$ and $(40~\mathrm{m},10~\mathrm{m})$, respectively. A circular service region with radius $10~\mathrm{m}$ and center located at $(40~\mathrm{m},0~\mathrm{m})$ is considered, where the users are randomly deployed within the coverage region. The wireless propagation environment is modeled using the planar far-field geometric channel model, with identical numbers of transmit-side and receive-side propagation paths, i.e., $K_t=K_r$. The angular parameters associated with the propagation paths, namely $\theta_i^{e}$, $\vartheta_j^{e}$, $\theta_i^{a}$, and $\vartheta_j^{a}$, are independently generated according to the uniform distribution on $[0,\pi]$, where $1\leq i\leq K_t$ and $1\leq j\leq K_r$. Unless otherwise stated, all numerical results are obtained by averaging over $10^4$ independently generated channel realizations. Further, the BS--RIS and RIS--user links are assumed to contain line-of-sight (LoS) components. Under this assumption, the diagonal entries of the BS--RIS path response matrix $\boldsymbol{\Omega}$ follow $[\boldsymbol{\Omega}]_{1,1}\sim\mathcal{CN}\left(0,\frac{\bar{\varkappa}}{\bar{\varkappa}+1}R_0\left(\frac{d}{d_0}\right)^{-\hat{\iota}}\right)$ and $[\boldsymbol{\Omega}]_{j,j}\sim\mathcal{CN}\left(0,\frac{1}{\bar{\varkappa}+1}R_0\left(\frac{d}{d_0}\right)^{-\hat{\iota}}/(K_t-1)\right)$ for $j=2,\ldots,K_t$, where $\bar{\varkappa}$ denotes the Rician factor and $R_0$ represents the average channel gain at the reference distance $d_0=1~\mathrm{m}$. The RIS--user channels $\mathbf{h}_{r,n}\in\mathbb C^{Q\times1}$ are modeled using Rician fading and are assumed to be independent of the MAs positions. With an appropriate RIS placement, the RIS-related links are assumed to exhibit approximately free-space attenuation with a path-loss exponent of
$\hat{\iota}=2.2$ \cite{pan2020multicell}. Unless specified otherwise, the following simulation parameters are adopted: ${\mu}_{n}^{2}=\mu \|{\mathbf{F}}_{n}\|_2$ with $\mu\in[0,1)$ \cite{zheng2023zero}, $\bar{\sigma}_n^2=\tilde{\sigma}_{\mathrm d}^2=-114~\mathrm{dBm}$, $\delta_n=\delta=0.05,\ \forall n\in\mathcal N$, $R_0=-30~\mathrm{dB}$, $R_{\min}=1~\mathrm{bps/Hz}$, $D_{\min}=\lambda/2$, $P_{\max}=50~\mathrm{dBm}$, $K_t=K_r=10$, $\bar{\varkappa}=3$, $ \widehat{\kappa}_n=-38~\mathrm{dBm}$, $N=3$, $\widehat{w}_n=6400$, $\widehat{y}_n=0.003$, and $\widehat{W}_n=0.2\,\mathrm{mW}$ \cite{asif2026robust,xiong2017rate}. In addition, the MAs deployment region is specified by $\mathcal Z_t=\left[-\frac{D}{2},\frac{D}{2}\right]\times\left[-\frac{D}{2},\frac{D}{2}\right]$, where $D=3\lambda$ and $\lambda=0.1~\mathrm{m}$ \cite{xiao2024multiuser}.

The performance of the proposed framework is assessed against the following four benchmark schemes. The proposed scheme is denoted by MAs-RIS-HIs throughout the numerical results. All benchmarks use the same channel realizations, CSI uncertainty model, and hardware-impairment parameters. 

\begin{itemize}
	\item \textbf{FPA-RIS-HIs:}
	This scheme fixes the BS antenna positions while optimizing the remaining design variables under the
	same constraints as the proposed scheme. This comparison evaluates the benefit of antenna-position optimization.
	
	\item \textbf{SDMA-RIS-HIs:}
	This scheme replaces RSMA with SDMA by disabling the common stream and transmitting only private streams. The remaining design variables are jointly optimized under the same system assumptions and constraints to assess the benefit of rate splitting.
	
	\item \textbf{PSO-RIS-HIs:}
	This scheme employs particle swarm optimization (PSO) \cite{ding2024movable} to jointly search over the design variables of the considered system. Each particle represents a candidate design and is updated based on its own best solution and the best solution found by the swarm. Candidate designs are evaluated using the same sum-rate objective and constraints.
	
	\item \textbf{CWS-RIS-HIs:}
    This scheme employs coordinate-wise search (CWS), inspired by the alternating position selection 
    benchmark in \cite{xiao2024multiuser}, and extends the search to all design variables. It evaluates 
    candidate changes to one coordinate at a time while keeping the others fixed and accepts feasible changes that improve the sum-rate. The step sizes are progressively reduced until the stopping criterion is met.
\end{itemize}
 
\begin{figure}[!t]
	\centering
	\includegraphics [width=0.36\textwidth]{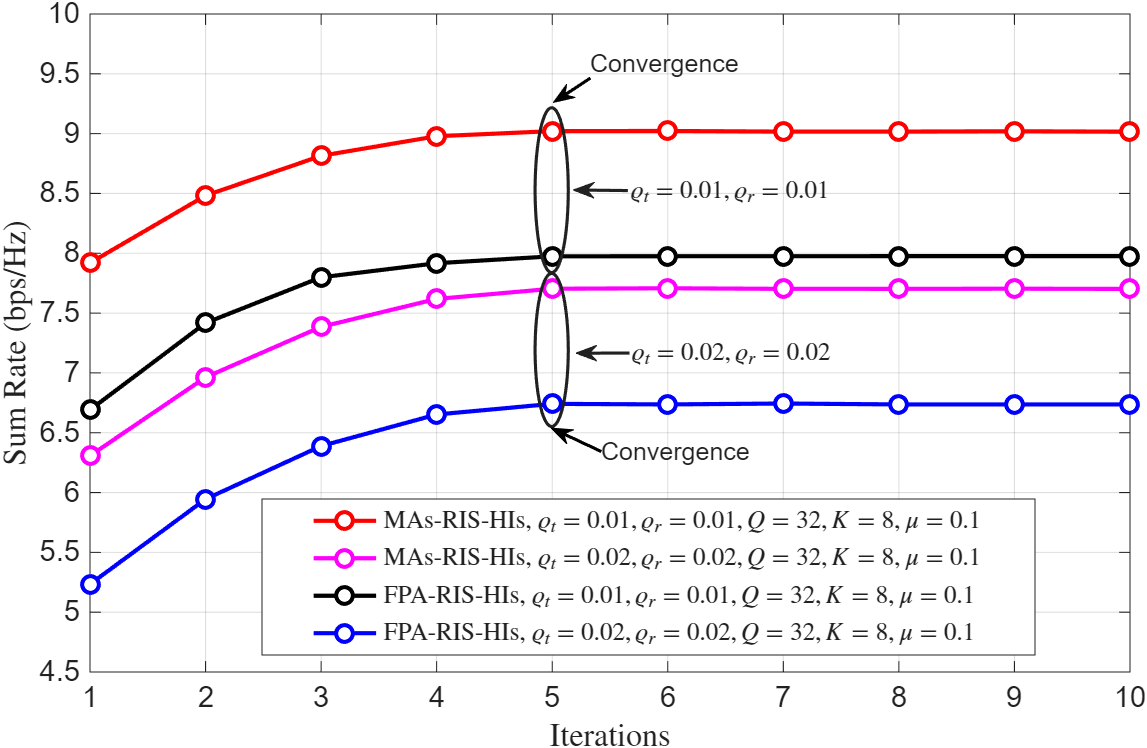}
	\caption{Convergence for different values of $\varrho_t$ and $\varrho_r$.}
	\label{f3}
\end{figure} 

\begin{figure}[t]
	\centering
	\includegraphics [width=0.36\textwidth]{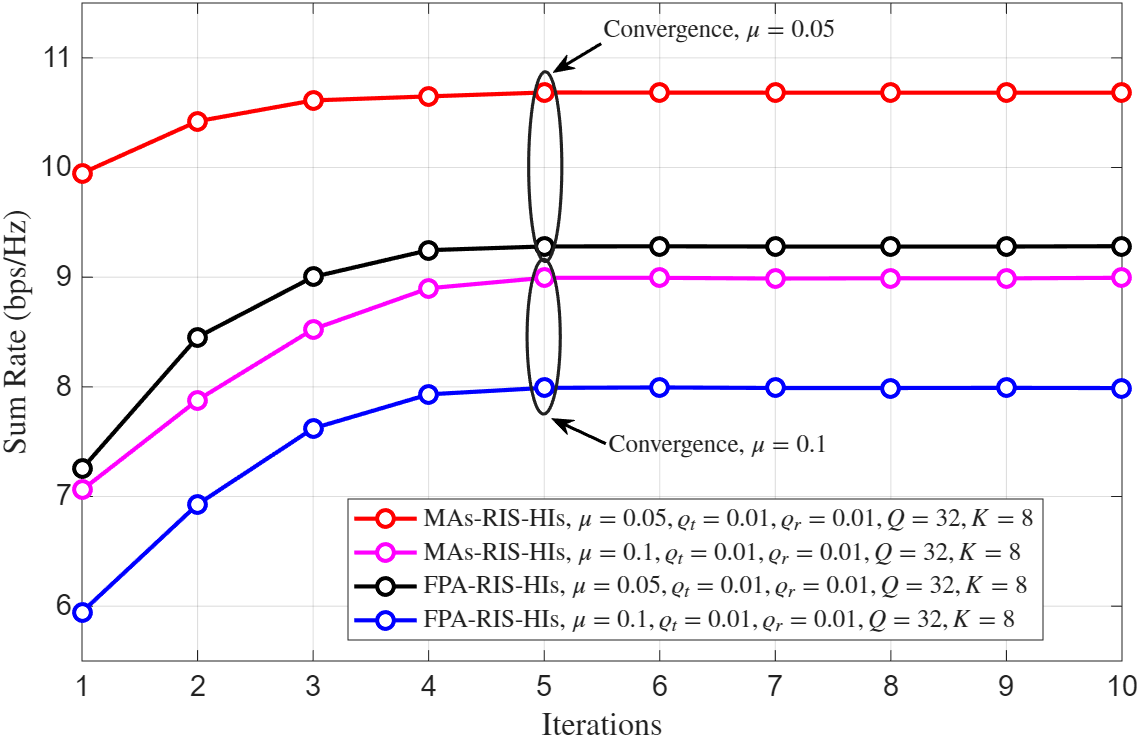}
	\caption{System convergence under different values of $\mu$.}
	\label{f4}
\end{figure}

  \begin{figure}[!t]
	\centering
	\includegraphics [width=0.36\textwidth]{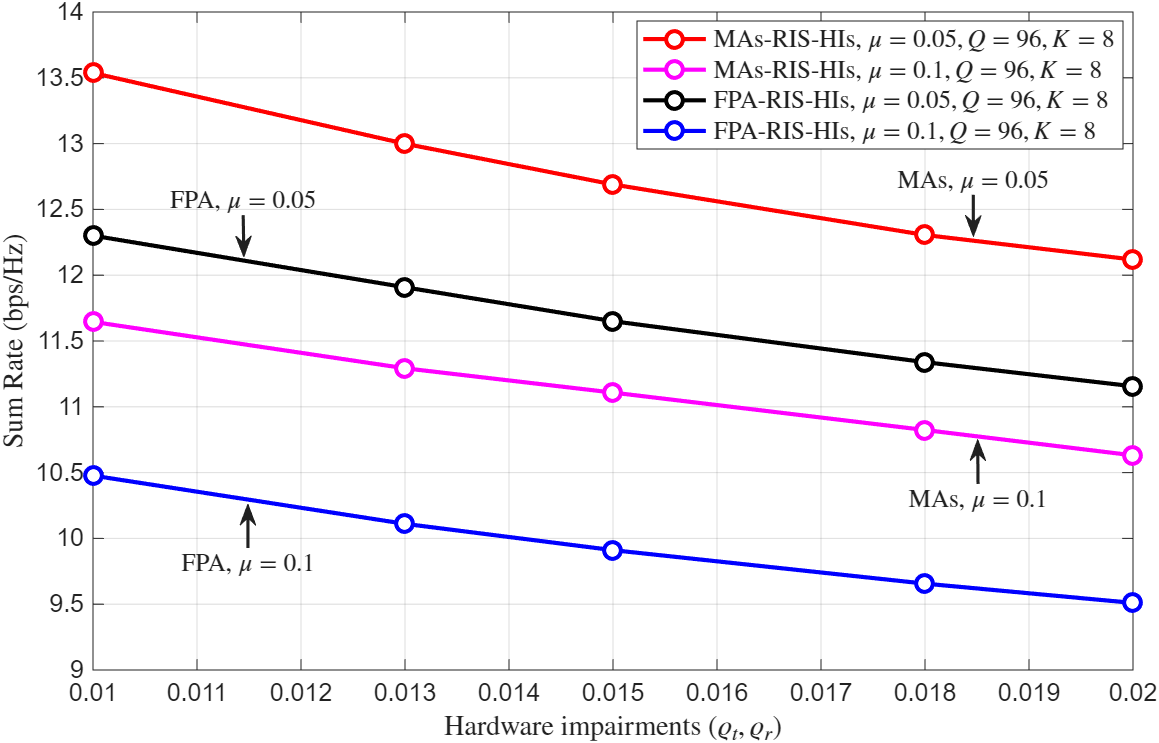}
	\caption{Performance under increasing transceiver hardware impairment levels for different values of $\mu$.}
	\label{f5}
\end{figure}

\begin{figure}[h]
	\centering
	\includegraphics [width=0.36\textwidth]{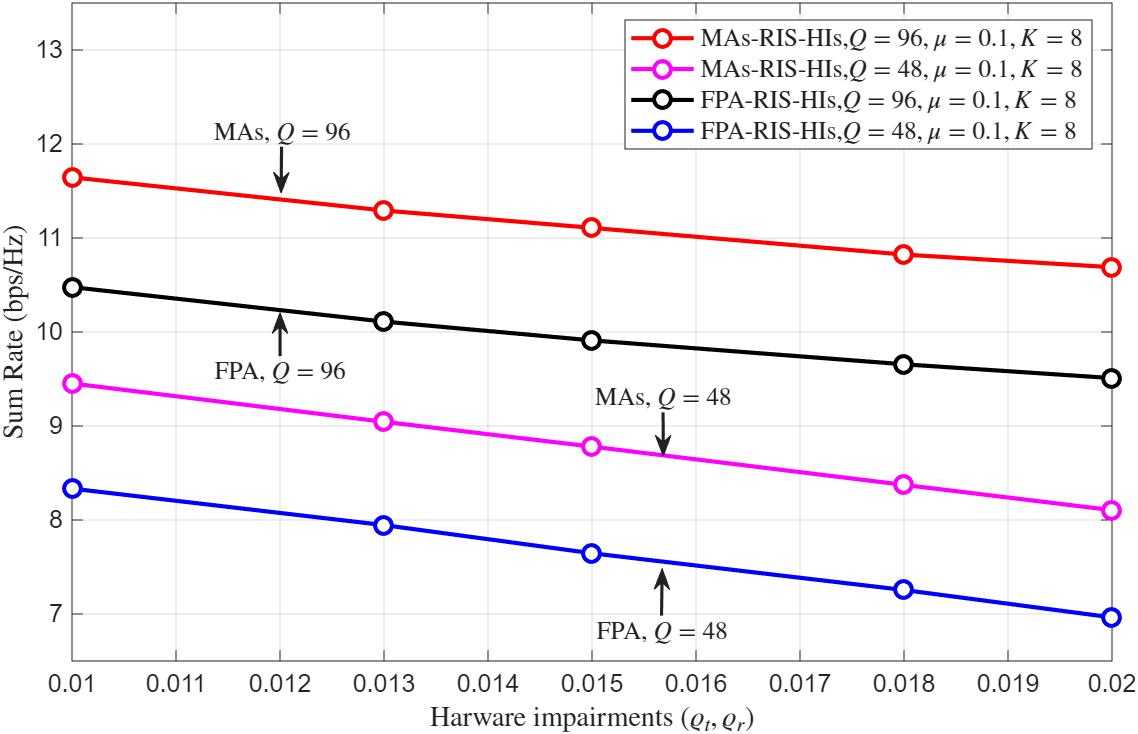}
	\caption{Impact of transceiver hardware impairment levels on the achievable sum-rate for different values of $Q$.}
	\label{f6}
\end{figure}

\begin{figure}[h]
\centering
\includegraphics [width=0.36\textwidth]{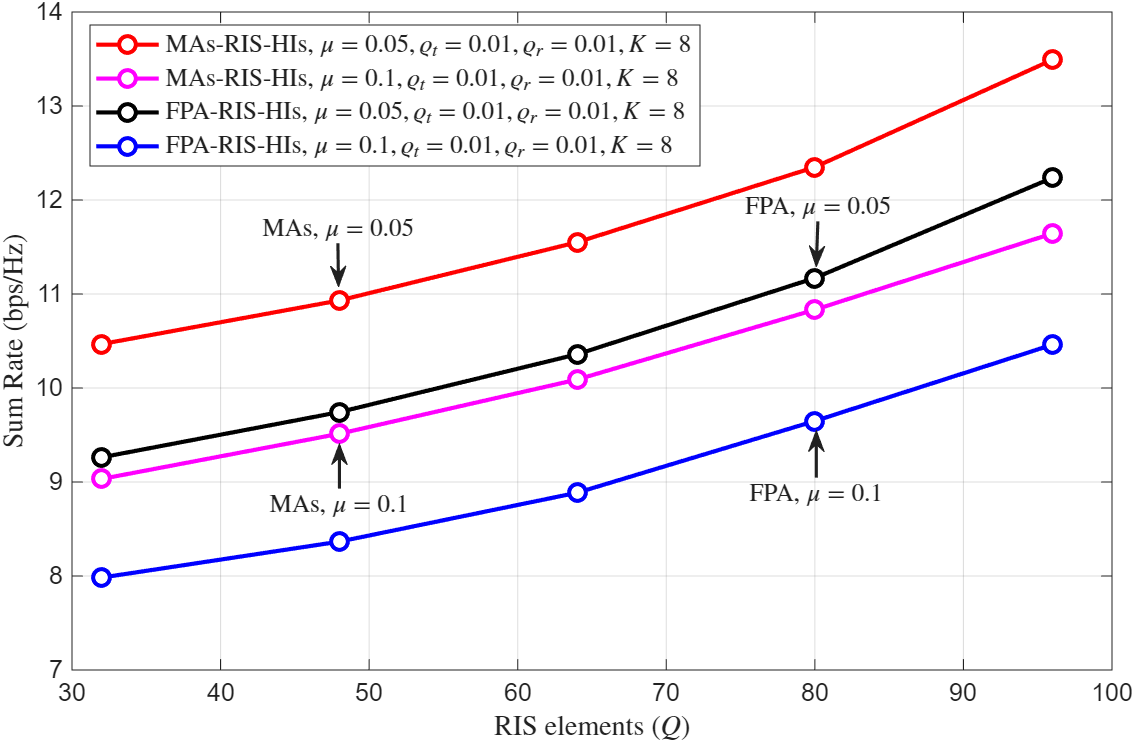}
\caption{Impact of RIS size $Q$ on the achievable system performance for different values of $\mu$.
}
\label{f7}
\end{figure}

\begin{figure}[h]
	\centering
	\includegraphics [width=0.36\textwidth]{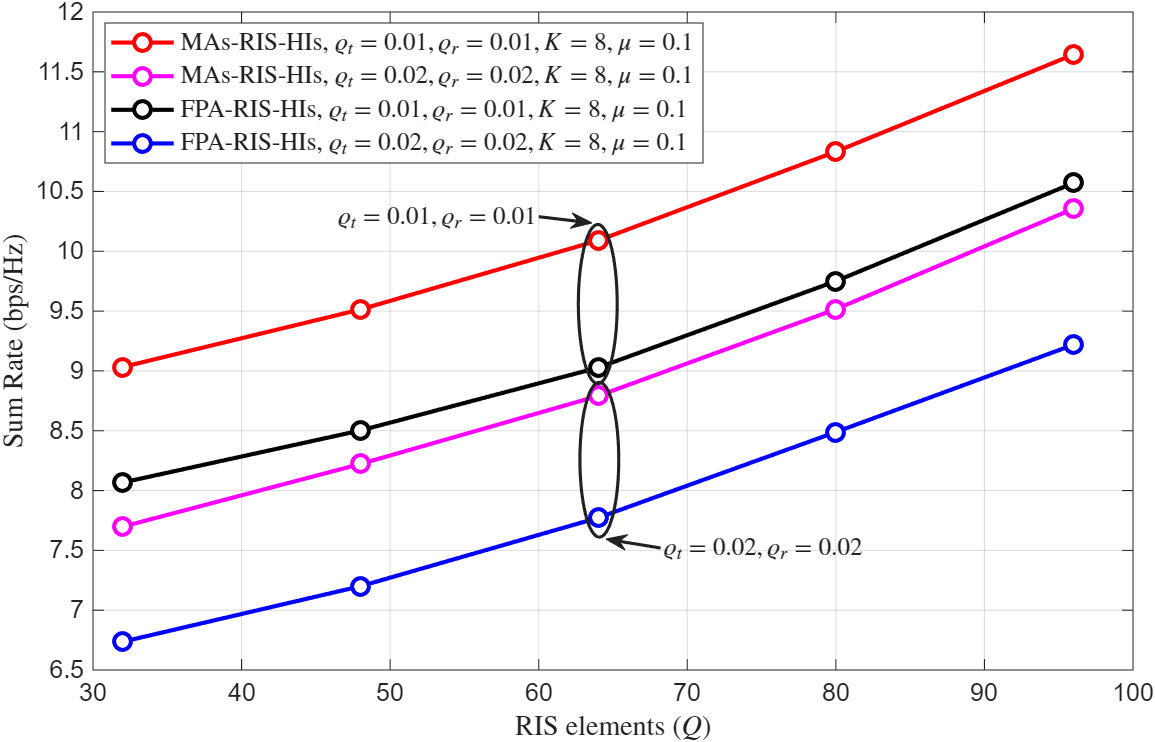}
	\caption{Achievable sum-rate versus RIS size $Q$ for different values of $\varrho_t$ and $\varrho_r$.}
	\label{f8}
\end{figure} 

\begin{figure}[h]
	\centering
	\includegraphics [width=0.36\textwidth]{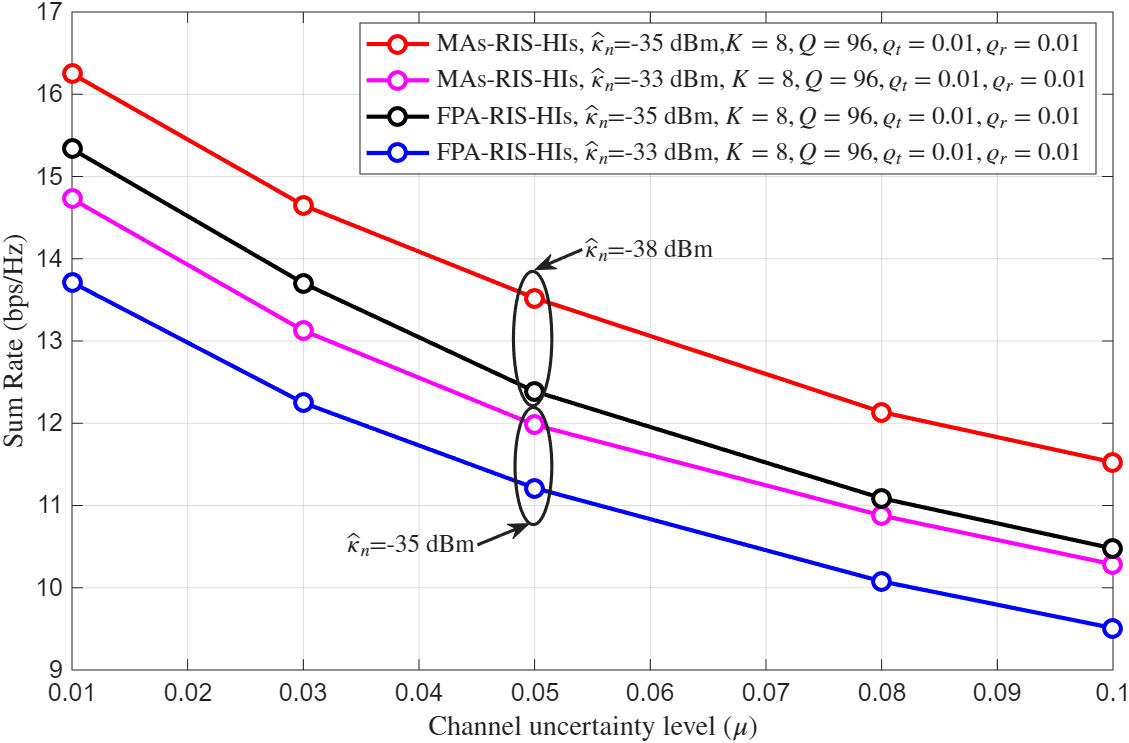}
	\caption{System performance versus channel uncertainty level $\mu$ for different values of $\widehat{\kappa}_n$.}
	\label{f9}
\end{figure} 

\begin{figure}[h]
	\centering
	\includegraphics [width=0.36\textwidth]{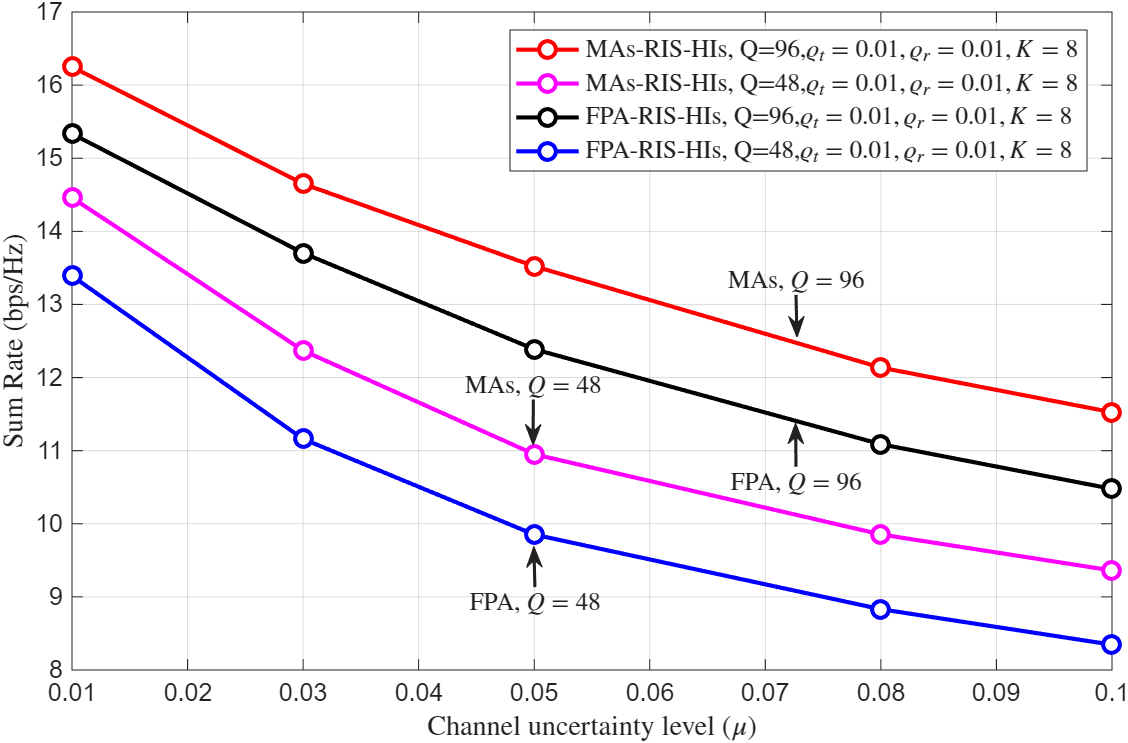}
	\caption{Impact of channel uncertainty level $\mu$ on system performance for different RIS sizes $Q$.}
	\label{f10}
\end{figure} 

\begin{figure}[h]
	\centering
	\includegraphics [width=0.36\textwidth]{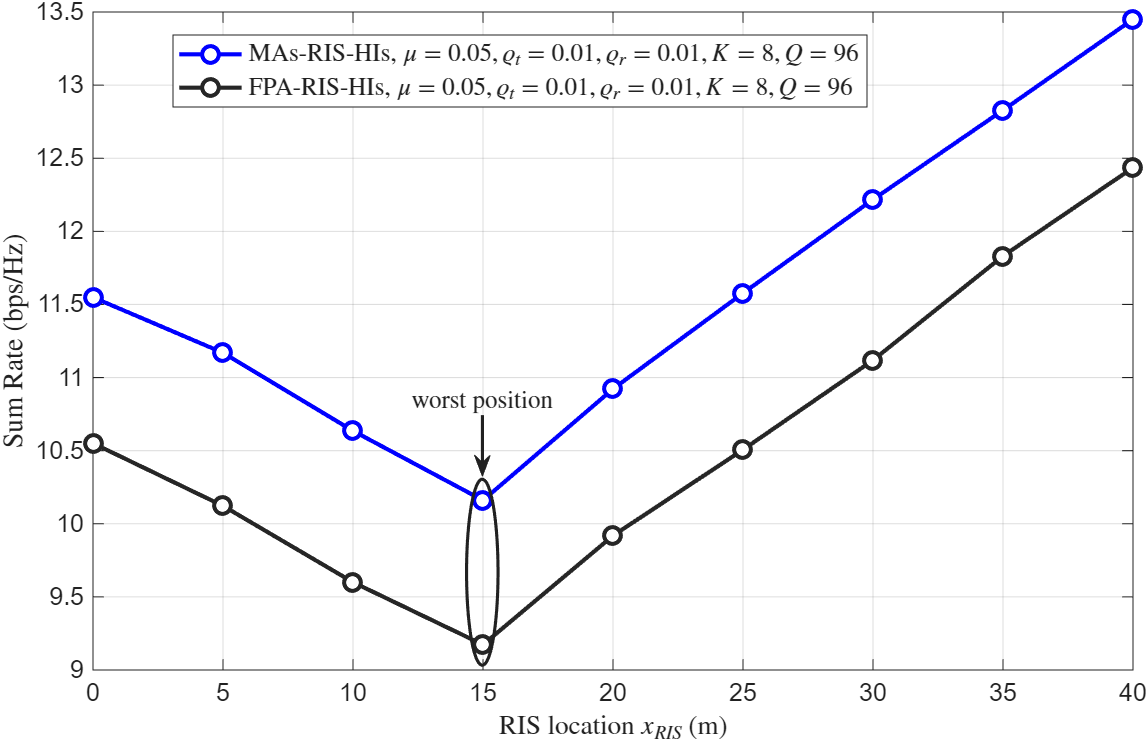}
	\caption{Impact of RIS placement on the achievable sum-rate.}
	\label{fig:ris_placement}
\end{figure} 

\begin{figure}[h]
	\centering
	\includegraphics [width=0.36\textwidth]{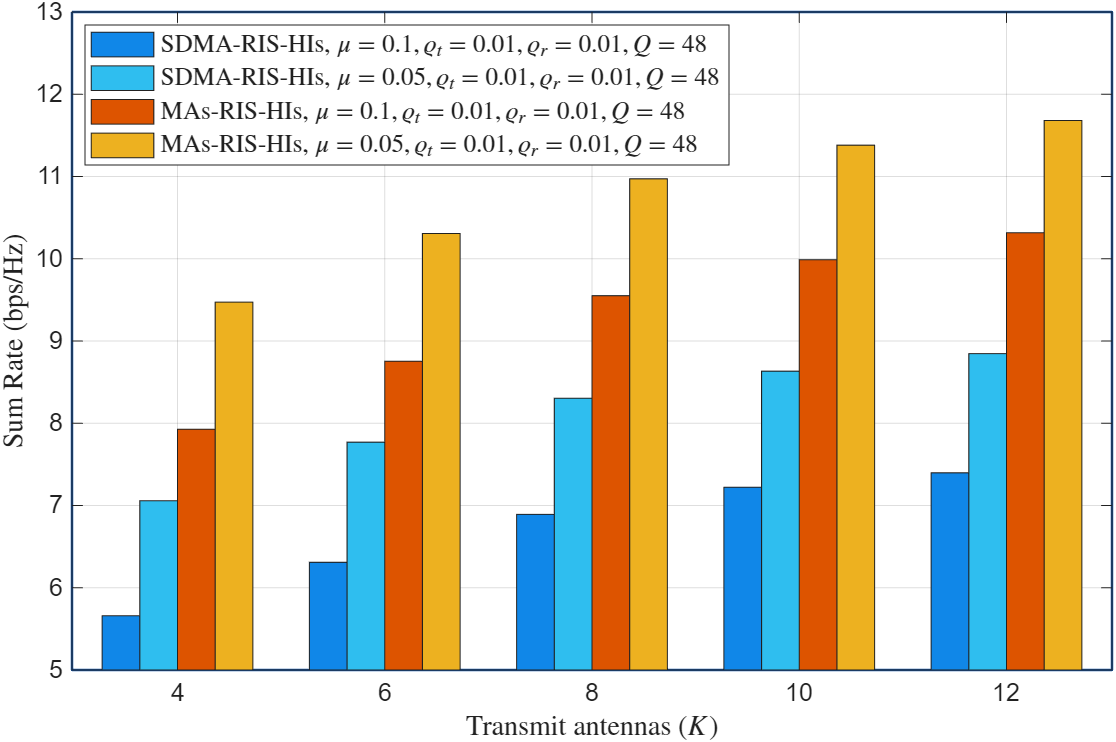}
	\caption{Comparative performance analysis under different numbers of MAs and values of $\mu$.}
	\label{f11}
\end{figure} 

\begin{figure}[h]
	\centering
	\includegraphics [width=0.36\textwidth]{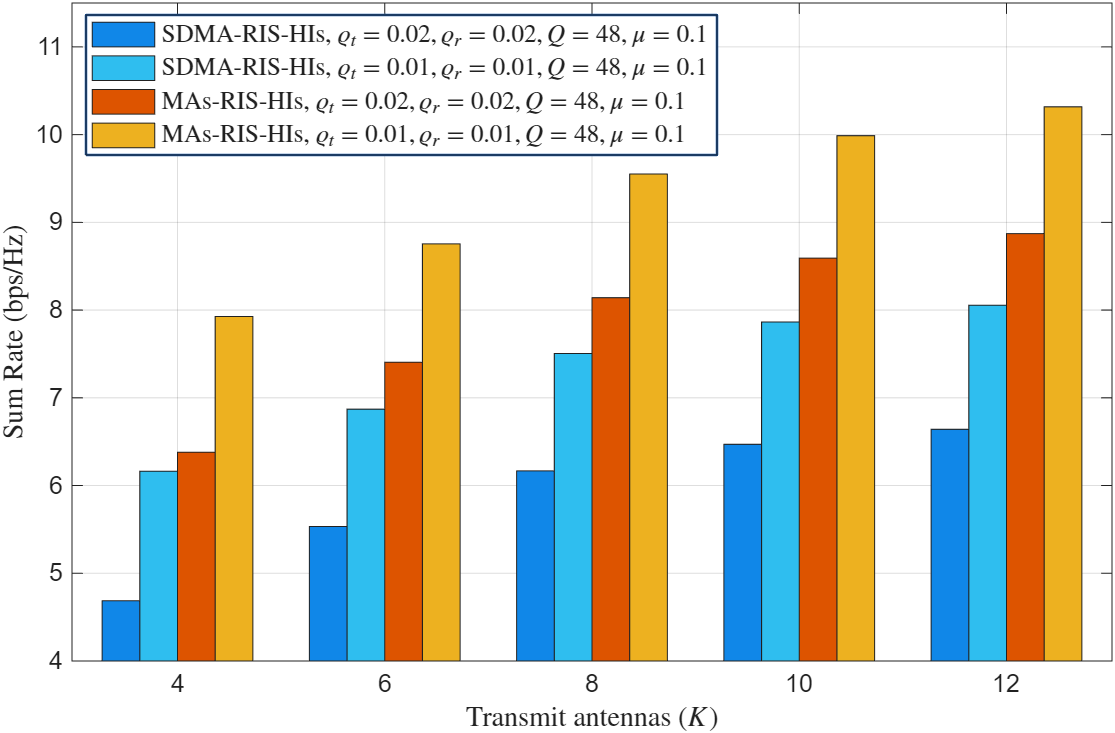}
	\caption{Comparative performance evaluation under different numbers of MAs and impairment levels.}
	\label{f12}
\end{figure} 

\begin{figure}[h]
	\centering
	\includegraphics [width=0.36\textwidth]{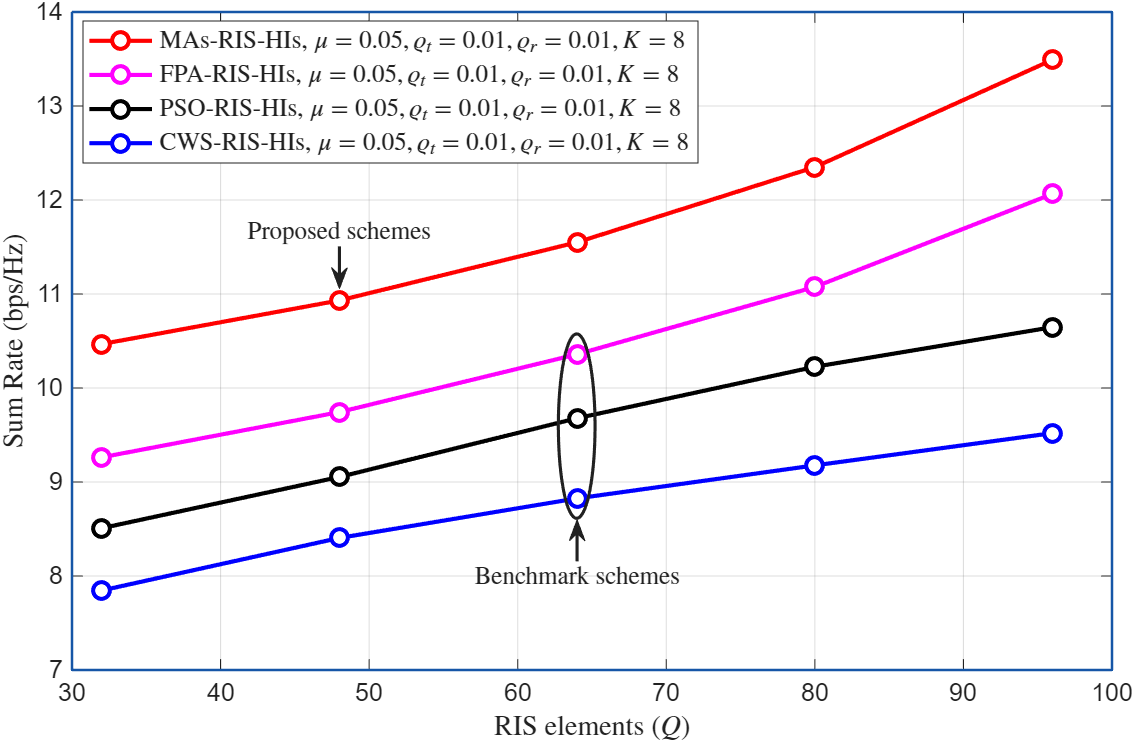}
	\caption{Sum-rate performance against benchmarks for varying RIS sizes.}
	\label{fig:benchmarks}
\end{figure}

The convergence characteristics of the proposed optimization framework, denoted by MAs-RIS-HIs, under different parameter settings are presented in Fig.~\ref{f3} and Fig.~\ref{f4}. In Fig.~\ref{f3}, the impact of transceiver hardware impairments is investigated by varying the transmitter and receiver impairment coefficients, $\varrho_t$ and $\varrho_r$. The results indicate that higher impairment levels lead to a noticeable reduction in the achievable sum-rate, which is attributed to the additional distortion introduced by hardware imperfections and the resulting degradation in the received SINR. Fig.~\ref{f4} shows the impact of varying channel uncertainty levels, represented by $\mu$, on the convergence behavior of the proposed framework. It is observed that the larger values of $\mu$ lead to lower achievable rates owing to the increased mismatch between the estimated and actual channel conditions. Despite these adverse effects, the proposed algorithm exhibits monotonic improvement in the achievable sum-rate and converges within a practical number of iterations for all considered parameter settings, thereby confirming its effectiveness and robustness.

Next, Fig.~\ref{f5} and Fig.~\ref{f6} examine the influence of transceiver hardware impairments on the achievable sum-rate under different channel uncertainty levels $\mu$ and numbers of RIS elements $Q$, respectively. Specifically, the transmitter and receiver impairment coefficients, $\varrho_t$ and $\varrho_r$, are varied to evaluate their impact on system performance. As the transceiver hardware impairment levels increase, a consistent reduction in the achievable sum-rate is observed across all considered values of $\mu$ and $Q$. This performance deterioration is mainly caused by the distortion noise introduced by hardware impairments, thereby degrading the effective SINR and reducing the achievable sum-rate.

Further, Fig.~\ref{f7} and Fig.~\ref{f8} illustrate the effect of the RIS size on the achievable system performance for different channel uncertainty levels $\mu$ and transceiver hardware impairment coefficients $\varrho_t$ and $\varrho_r$, respectively. It can be observed that the achievable sum-rate increases steadily with the number of RIS reflecting elements $Q$ under all considered parameter settings. This performance improvement stems from the enhanced passive beamforming capability and the increased array gain offered by a larger RIS, which strengthens the effective cascaded channel. Since these rates are achieved subject to the users' energy-harvesting requirements, the results illustrate how additional RIS resources support higher information rates while maintaining the required energy delivery. Furthermore, the achievable sum-rate is adversely affected by larger values of $\mu$, $\varrho_t$, and $\varrho_r$. This behavior can be attributed to the combined impact of channel uncertainty and hardware-induced distortion, thereby degrading the effective SINR and reducing the achievable sum-rate performance.

The effect of channel uncertainty on the achievable system performance is investigated in Fig.~\ref{f9} and Fig.~\ref{f10} for different EH thresholds $\widehat{\kappa}_n$ and numbers of RIS elements $Q$, respectively. The results in both figures indicate that the achievable sum-rate gradually deteriorates as the channel uncertainty parameter $\mu$ increases. This performance degradation is caused by the growing discrepancy between the estimated and actual channel conditions, which limits the effectiveness of resource allocation and beamforming optimization. Specifically, Fig.~\ref{f9} shows that the achievable sum-rate decreases with increasing values of the minimum harvested energy requirement $\widehat{\kappa}_n$. This is because a more stringent EH requirement allocates a larger fraction of the received signal power to energy harvesting, leaving less power available for information decoding and consequently reducing the achievable sum-rate. Moreover, the results presented in Fig.~\ref{f10} demonstrate that increasing the number of RIS elements improves the achievable sum-rate by providing higher passive beamforming gains. Nevertheless, the performance degradation caused by channel uncertainty remains evident across all considered system configurations.

Fig.~\ref{fig:ris_placement} shows the achievable sum-rate as the horizontal RIS coordinate \(x_{\mathrm{RIS}}\) varies from \(0\) to \(40\) m. For both the proposed MAs-RIS-HIs scheme and the FPA-RIS-HIs benchmark, the sum-rate first decreases, reaching its lowest value among the tested locations at \(x_{\mathrm{RIS}}=15\) m, and then increases as the RIS moves toward the user region. This behavior reflects the combined attenuation of the BS–RIS and RIS–user links. Moving the RIS closer to the users shortens the RIS–user distances but increases the BS–RIS distance, resulting in a trade-off between the two hops. In the considered setup, both schemes achieve their highest sum-rate at \(x_{\mathrm{RIS}}=40\) m, indicating that deployment near the user region is more favorable within the tested range. The proposed MAs-RIS-HIs scheme also maintains a higher sum-rate than FPA-RIS-HIs at every tested location, showing that antenna-position optimization remains beneficial across different RIS placements.

Fig.~\ref{f11} and Fig.~\ref{f12} compare the performance of the proposed framework with its SDMA-based benchmark, denoted by SDMA-RIS-HIs, under increasing numbers of transmit MAs. Specifically, Fig.~\ref{f11} and Fig.~\ref{f12} consider different channel uncertainty levels $\mu$ and transceiver hardware impairment coefficients $\varrho_t$ and $\varrho_r$, respectively. Simulation results depict that the proposed framework achieves higher sum-rate performance than the SDMA benchmark under all examined settings of $K$, $\mu$, $\varrho_t$, and $\varrho_r$. Both schemes satisfy the same energy-harvesting requirements, so this comparison demonstrates the benefit of RSMA in improving information transmission under the considered SWIPT constraints. Common-stream transmission provides additional flexibility in managing inter-user interference while contributing to harvested energy, since its cancellation occurs only in the information-decoding branch. In addition, the achievable sum-rate increases monotonically as the number of transmit MAs grows, regardless of the considered system configuration. This performance gain is attributed to the additional spatial degrees of freedom provided by the increased number of transmit MAs, which offer greater flexibility in antenna placement and facilitate more effective beamforming. Consequently, the transmit antennas can be positioned at more favorable positions, resulting in a significant improvement in the achievable system performance.

Finally, Fig.~\ref{fig:benchmarks} compares the sum-rate of the proposed MAs-RIS-HIs scheme with FPA-RIS-HIs, PSO-RIS-HIs, and CWS-RIS-HIs for different numbers of RIS elements. The sum-rate increases for all four schemes, as additional reflecting elements provide greater flexibility to strengthen the desired signals and manage inter-user interference. The performance gaps between the proposed scheme and the PSO-RIS-HIs and CWS-RIS-HIs benchmarks widen as the number of RIS elements increases, indicating that the proposed algorithm uses the additional design freedom more effectively under the considered simulation settings. It is worth mentioning that MAs-RIS-HIs outperforms FPA-RIS-HIs, which in turn achieves higher sum-rates than PSO-RIS-HIs and CWS-RIS-HIs despite using fixed antennas. These results highlight both the benefit of antenna-position optimization and the effectiveness of the proposed optimization approach.

\section{Conclusion}
This paper has proposed a robust transmission design for an RIS-enabled MAs-assisted multi-user RSMA-SWIPT system under CSI uncertainty and residual HIs. By jointly optimizing common-rate allocation, transmit beamforming, RIS reflection coefficients, power-splitting ratios, and MAs positions, a robust sum-rate maximization framework was developed while incorporating a practical non-linear energy harvesting model. To efficiently solve the resulting highly coupled non-convex problem, an iterative optimization framework was developed by successively decomposing the original problem into active beamforming, RIS reflection design, power-splitting ratio, and MAs position optimization subproblems. Numerical results demonstrated that the proposed design achieved significant sum-rate improvements and enhanced robustness under CSI uncertainty and residual HIs while maintaining fast convergence and stable performance. These findings highlight the effectiveness of the proposed framework for future RIS-assisted MAs-enabled RSMA-SWIPT networks operating under practical channel and hardware conditions.

\bibliographystyle{IEEEtran}
\bibliography{Ref}
\vskip -2\baselineskip plus -2fil

\end{document}